\documentclass[smallextended]{svjour3}       % onecolumn (second format)
\smartqed  % flush right qed marks, e.g. at end of proof
\usepackage{graphicx}
\journalname{Empirical Software Engineering}

\usepackage{hyperref}
\usepackage[utf8]{inputenc} 
\usepackage[T1]{fontenc}
\usepackage{microtype}

\usepackage{booktabs, multirow} % for borders and merged ranges
\usepackage{soul}% for underlines

\usepackage{color}
\usepackage{pdfcomment}
\usepackage{todonotes}

\usepackage{listings}
\usepackage{color}
\definecolor{light-grey}{RGB}{225, 225, 225}
\usepackage[round]{natbib}
\usepackage[nocompress]{cite}

\usepackage{algorithm}
\usepackage{algpseudocode}
\usepackage{float}

\usepackage{xpatch}
\xpatchcmd\algorithmic
  {\labelwidth 1.2em}
  {\labelwidth .7em}
  {}{\fail}

\algrenewcommand\algorithmicindent{0.8em}%

\usepackage{amssymb}

\usepackage{subfig}
\usepackage{makecell}

\newcommand{\stateafl}[1]{\textsc{StateAFL}}
\newcommand{\aflnet}[1]{\textsc{AFLnet}}
\newcommand{\aflnwe}[1]{\textsc{AFLnwe}}
\newcommand{\afl}[1]{\textsc{AFL}}

\begin{document}

\title{\stateafl{}: Greybox Fuzzing for Stateful Network Servers}

\author{Roberto Natella}

\institute{R. Natella \at
              Università degli Studi di Napoli Federico II, Via Claudio 21, 80125, Napoli, Italy\\
              \email{roberto.natella@unina.it}
}

\date{}

\maketitle

\begin{abstract}
Fuzzing network servers is a technical challenge, since the behavior of the target server depends on its state over a sequence of multiple messages. Existing solutions are costly and difficult to use, as they rely on manually-customized artifacts such as protocol models, protocol parsers, and learning frameworks. The aim of this work is to develop a greybox fuzzer (\stateafl{}) for network servers that only relies on lightweight analysis of the target program, with no manual customization, in a similar way to what the \afl{} fuzzer achieved for stateless programs. The proposed fuzzer instruments the target server at compile-time, to insert probes on memory allocations and network I/O operations. At run-time, it infers the current protocol state of the target server by taking snapshots of long-lived memory areas, 
and by applying a fuzzy hashing algorithm (Locality-Sensitive Hashing) to map memory contents to a unique state identifier.
The fuzzer incrementally builds a protocol state machine for guiding fuzzing. 

We implemented and released \stateafl{} as open-source software. As a basis for reproducible experimentation, we integrated \stateafl{} with a large set of network servers for popular protocols, with no manual customization to accomodate for the protocol. The experimental results show that the fuzzer can be applied with no manual customization on a large set of network servers for popular protocols, and that it can achieve comparable, or even better code coverage and bug detection than customized fuzzing. Moreover, our qualitative analysis shows that states inferred from memory better reflect the server behavior than only using response codes from messages.
\keywords{Security \and Fuzzing \and Network Servers}
%\PACS{PACS code1 \and PACS code2 \and more}
%\subclass{MSC code1 \and MSC code2 \and more}
\end{abstract}

\section{Introduction}

According to recent statistics \citep{google0day,mit2021attacks}, high-severity software vulnerabilities of network servers have been on the rise, and will likely still be in the near future. Network servers are a critical part of the attack surface of IT infrastructures, as they are openly exposed to malicious users over local networks and the Internet, and can be attacked with malformed traffic to cause a denial-of-service (e.g., crashing the server), and to execute arbitrary code on the server machine to perpetrate further attacks. For this reason, any vulnerability not yet found by developers (\emph{0-days}) has a significant economic value for attackers \citep{guo2021revenue}.

\emph{Fuzzing} is a relevant security testing technique to identify such vulnerabilities, by automatically generating large volumes of malformed inputs. However, fuzzing network servers is still a technical challenge, since the input space of network servers is strictly regulated by a \emph{stateful protocol}. Therefore, the behavior of the server, and its vulnerabilities, depend on a sequence of several messages exchanged over time, which determine the \emph{state} of the server. 
Examples of well-know stateful protocols include cryptographic ones such as TLS \citep{de2015protocol,fiterau2020analysis}, file transfer and messaging protocols such as FTP, SMB, and SMTP \citep{antunes2011automatically,comparetti2009prospex}, and multimedia protocols such as SIP \citep{banks2006snooze,alrahem2007interstate} and RSTP \citep{pham2020aflnet}. All of these protocols are selective with respect to which messages they can receive at a given time, and which actions they can perform, depending on previous messages in a session.

The existing stateful protocol fuzzing techniques and tools can only be applied with a significant effort, which has prevented their widespread adoption so far: \emph{generation-based} fuzzers require formal specifications manually written by human experts, based on their detailed knowledge of the protocols  \citep{beSTORM,Defensics,Metasploit}; \emph{learning-based} fuzzers infer the protocol state machine at a significant computational cost, and still require custom implementations of wrappers to abstract protocol messages in efficient ways \citep{de2015protocol,fiterau2020analysis}.

\emph{Coverage-driven fuzzing} techniques have recently emerged as a popular solution, as demonstrated by the widespread adoption of the \afl{} fuzzer and similar tools \citep{afl,metzman2021fuzzbench,manes2019art,boehme2021fuzzing}. 
For example, as of June 2021, OSS-Fuzz has found over $30,000$ bugs in $500$ open source projects \citep{oss-fuzz-github,serebryany2017oss}, with more and more open-source projects being integrated by the community \citep{adalogics}. 
This success could only be possible thanks to its fully-automated approach, which is based on unsupervised evolution of fuzz inputs, using simple and robust heuristics. 
However, research on coverage-driven fuzzing for stateful protocols is still at an early stage \citep{pham2020aflnet,feng2021snipuzz}. These recent approaches infer protocol states by analyzing the contents of messages (e.g., \emph{status codes}), using message parsers that are specifically developed for the protocol under test. Moreover, it is difficult for these approaches to fuzz many protocols, which only embed little or no state information within messages. These problems are a limiting factor towards securing more stateful network servers through fuzzing.

In this work, we propose a new solution for \emph{stateful coverage-driven} fuzzing (\stateafl{}). Similarly to coverage-driven fuzzing, we inject code in the target binary using compile-time instrumentation techniques. The injected code infers protocol state information by: tracking memory allocations and network I/O operations; at each request-reply exchange, taking snapshots of long-lived memory areas; 
and applying fuzzy hashing (Locality-Sensitive Hashing, LSH) to map each in-memory state to a unique protocol state identifier.
This approach does not rely on state information from network messages, and does not require developers to implement custom message parsers for extracting such state information. 
The aim of this approach is to contribute towards a completely-automated solution for stateful protocol fuzzing, similarly to what \afl{} was able to achieve for stateless programs, in order to promote a wider application of fuzzing in real-world systems. We note that fuzzing research achieved significant progress from the point of view of fuzzing algorithms, but we are still witnessing at critical vulnerabilities (e.g., the well-known case of Hearthbleed \citep{wheeler2020how}) that in hindsight could have been easily prevented with fuzzing. Moreover, empirical research also showed that fuzzing a new system for the first time is likely to find security bugs \citep{bohme2020fuzzing}. For these reasons, it is now a priority to make fuzzing more broadly applicable, as it is still too difficult to setup fuzzing to target new systems. 
In the case of stateful network fuzzing, StateAFL overcomes the issues of writing custom parsers to extract individual requests from seed inputs, and to extract status codes from response messages from the target server. These issues make fuzzing less accessible for developers that are new to this technique, since they are not inclined to write more code to use a fuzzing tool unfamiliar to them. Moreover, StateAFL is even applicable for protocols that do not provide any explicit status code in the messages, such as in the TLS protocol in our experiments, or where the status code only represents the status of the last request executed by the server instead of the protocol state, as in FTP and HTTP.

To assess the feasibility of the approach, we implemented and publicly released \stateafl{} as open-source software. Moreover, to support reproducible experimentation, we integrated \stateafl{} with a publicly-available benchmark of 13 open-source network servers, the largest experimental setup among stateful network fuzzing studies to the best of our knowledge. Our proposed approach allowed us to integrate \stateafl{} with no manual customization of the fuzzer to accomodate for the protocols under test. 
The experimental evaluation shows that \stateafl{} is a robust approach that can be applied to diverse network servers without requiring any protocol customization. Moreover, \stateafl{} can achieve comparable, or even better code coverage and bug detection than previous solutions based on stateless coverage-driven fuzzing and on stateful, protocol-customized fuzzing. 
We also qualitatively analyze state information both from parsing response codes returned by the target server, and from inference based on long-lived data. We found that using response codes provides misleading representation of the protocol state, leading to redundant states in the inferred protocol state machine and wasted fuzz inputs.

In summary, this paper presents the following contributions:
\begin{itemize}

\item A novel coverage-driven strategy for fuzzing stateful network servers, based on compile-time instrumentation and fuzzy hashing techniques to automatically infer protocol states from process memory;

\item An open-source fuzzing tool based on the proposed approach, available at \url{https://github.com/stateafl/stateafl}. Similarly to \afl{}, this fuzzer is designed to be applicable to a wide variety of targets without requiring customizations.

\item The integration of \stateafl{} in a public benchmark of network servers, with scripts to automate reproducible experimentation, available at \url{https://github.com/profuzzbench/profuzzbench}. 

\item An experimental evaluation of \stateafl{}, with respect to code coverage, bugs, and performance, along with a qualitative analysis of the inferred protocol states.

\end{itemize}

The paper is structured as follows. Section~\ref{sec:related} discusses related work on stateful fuzzing. Section~\ref{sec:proposed} presents the design and implementation of \stateafl{}. Section~\ref{sec:evaluation_plan} presents the experimental plan, and Section~\ref{sec:evaluation_results} presents the experimental results. Section~\ref{sec:conclusion} concludes the paper.

\section{Related work}
\label{sec:related}

\emph{Generation-based} fuzzers address stateful protocols by generating fuzz inputs using a \emph{model} of the protocol, to be provided by a human analyst \citep{beSTORM,Defensics,Metasploit}. The model specifies both the format of protocol messages (e.g., field types, message separators, etc.) and their sequencing over a session \citep{poll2015protocol}, typically in the form of a graph, such as finite state machines, prefix acceptor trees, and Markov chains. The completeness of the model is critical for the effectiveness of fuzzing, but it can be difficult to achieve, since protocol specifications (which are typically written in natural language) are prone to misinterpretations and costly to analyze, and do not cover proprietary protocol extensions \citep{antunes2011automatically}.

Several \emph{model learning} techniques have been proposed to compensate for these issues, by (semi-)automatically inferring the types and formats of messages, and protocol state machines. \emph{Passive} learning techniques infer from a corpus of network traces, using sequence alignment techniques (e.g., the Needleman-Wunsch algorithm) and statistical techniques (e.g., clustering into message types, and correlation of message fields) \citep{duchene2018state,kleber2018nemesys}. \emph{Active} learning techniques interact with the protocol server during the learning process, in order to refine the model and to elicit new protocol behaviors (e.g., based on Angluin's $L*$ algorithm and derivates) \citep{de2015protocol,fiterau2020analysis}. Both passive and active learning techniques provide valuable support for the human analyst, but cannot fully automate the process. For example, active learning can suffer from convergence issues and are applicable to finite input alphabets of modest size; thus, it needs an ad-hoc \emph{mapper} to abstract protocol messages from/to the learner, to be tailored for the system-under-test (e.g., TLS-Attacker for the TLS protocol) \citep{somorovsky2016systematic}. More powerful solutions leverage static and dynamic binary analysis (e.g., taint propagation analysis) to achieve full automation \citep{comparetti2009prospex,caballero2007polyglot}, but in practice these solutions are difficult to implement and to port across different systems, which limits their adoption \citep{harman2018start}.

\emph{Coverage-driven fuzzing} techniques have been adopted by \afl{}, \textsc{libFuzzer}, and other derivative tools \citep{manes2019art} as a more practical and automated solution. This form of fuzzing only relies on lightweight metrics collected from the target system at run-time (e.g., about code blocks and branches covered by the fuzz inputs), and iteratively mutates the fuzz inputs to maximize these metrics. Therefore, the fuzzer can start from an initial set of fuzz inputs (i.e., a \emph{seed} corpus) to automatically evolve them, without any a-priori knowledge about the protocol.

Only recently, coverage-driven fuzzing has been investigated for stateful protocols. \aflnet{} \citep{pham2020aflnet} extended \afl{} for fuzzing network protocols, by: structuring fuzz inputs into messages and applying mutation operators at message-level (e.g., by corrupting, dropping or injecting individual messages in a session); by learning a protocol state machine, where states are represented by response codes from the system-under-test; and by using the protocol state machine to prioritize mutations. \textsc{Snipuzz} \citep{feng2021snipuzz} tailored coverage-driven fuzzing to IoT protocols, where the system-under-test could not be instrumented to collect coverage information, because of lack of access to the firmware. Thus, \textsc{Snipuzz} also analyzes response codes, using them as indicators to identify sensitive bytes of the inputs (snippets) that trigger different paths in the target. 

This paper proposes a new approach for stateful protocol fuzzing. Our approach infers a protocol state machine on the basis on richer feedback than traditional coverage-driven fuzzing. The approach is not limited to analyze response codes, since response codes may provide a poor indication of the current state of the server. For example, in an HTTP-based protocol, successful GET and POST requests may both receive the same response code ($200$), but POST requests may have side-effects on the state of the server, which are not reflected in the response code. Moreover, the protocol may lack response codes, such as in the case of TLS, thus leaving the fuzzer without any guidance about the current protocol state. Finally, even when response codes available, the fuzzer must be tailored for the target protocol, in order to extract and parse response codes from the response messages. For these reasons, our approach does not rely on response codes, but adopts compile-time instrumentation to get more information from the system-under-test and to infer the current protocol state. Moreover, the proposed approach relieves the user from providing custom message parsers.

\section{Proposed approach}
\label{sec:proposed}

%\subsection{Overview}

We designed \stateafl{} to drive fuzzing based on \emph{protocol states} covered during executions. In general terms, a protocol state guides the behavior of a process, by defining \emph{which actions the process is allowed to take, which events it expects to happen, and how it will respond to those events} \citep{holzmann1991design}. For example, most Internet protocols standardize the protocol states and their transitions in \emph{Request for Comments} (RFC) documents, by describing them using prose in natural language or, in few cases, using finite state machines. Covering protocol states is a prerequisite for deeper code coverage of a protocol implementation, as some of its parts are only executed when the protocol reaches specific states. Moreover, exploring the protocol state space can uncover unintended or spurious behaviors of the protocol implementation that deviate from the protocol specification \citep{poll2015protocol}.

The \stateafl{} approach is designed around the fundamental receive-process-reply loop implemented by network servers. In this scheme, two parties (e.g., a \emph{client} and a \emph{server}) establish a \emph{session}, which consists of a series of \emph{request messages} and their corresponding \emph{reply messages} \citep{poll2015protocol}. As the session progresses, the current protocol state is updated accordingly. The fundamental loop can be summarized by the following simplified pseudo-code:

\vspace{0.5cm}
{
%\small
%\begin{algorithm}[t]
%\caption{\textsc{Whatever}}
\begin{algorithmic}%[1]

\State {$long$-$lived$ $data$ $\gets$ allocate()}
\While {iterate indefinitely}

    \State {$short$-$lived$ $data$ $\gets$ allocate()}
    \State {$request$ $\gets$ receive()}
    
    \State {$reply$ $\gets$ process($request$, $long$-$lived$ $data$, $short$-$lived$ $data$)}

    \State {send($reply$)}
    
    \State {deallocate($short$-$lived$ $data$)}
\EndWhile
\State {deallocate($long$-$lived$ $data$)}
\end{algorithmic}
%\end{algorithm}
}

The key idea of \stateafl{} is to infer the current protocol state by inspecting the contents of process memory at each iteration of this loop. The current protocol state is necessarily stored into data structures, such as in heap and stack memory, which are updated at each request-reply exchange. 
In particular, the protocol state is represented by \emph{long-lived data}, whose lifetime goes beyond an individual request-reply exchange, and spans across an entire session. Examples of such data are the current authentication status of a client, the current working directory, and enqueued inputs to be processed \citep{profuzzbench2021}. 
Conversely, \emph{short-lived data} have a short lifetime, as they store data only needed by one or few request-reply exchanges (such as, a buffer that temporarily holds the reply message). 
\stateafl{} follows the evolution of long-lived data structures thorough a session, and discards short-lived data. When fuzzing succeeds at reaching a new protocol state, the new state results in new contents of the long-lived data structures. Thus, the proposed approach takes a \emph{snapshot} of such data at the end of each request-reply exchange. Then, it uses this snapshot as a proxy for the current protocol state, by assigning a unique state identifier to each unique memory state through fuzzy hashing.

\begin{figure*}[!ht]
  \begin{center}
  \includegraphics[width=\columnwidth]{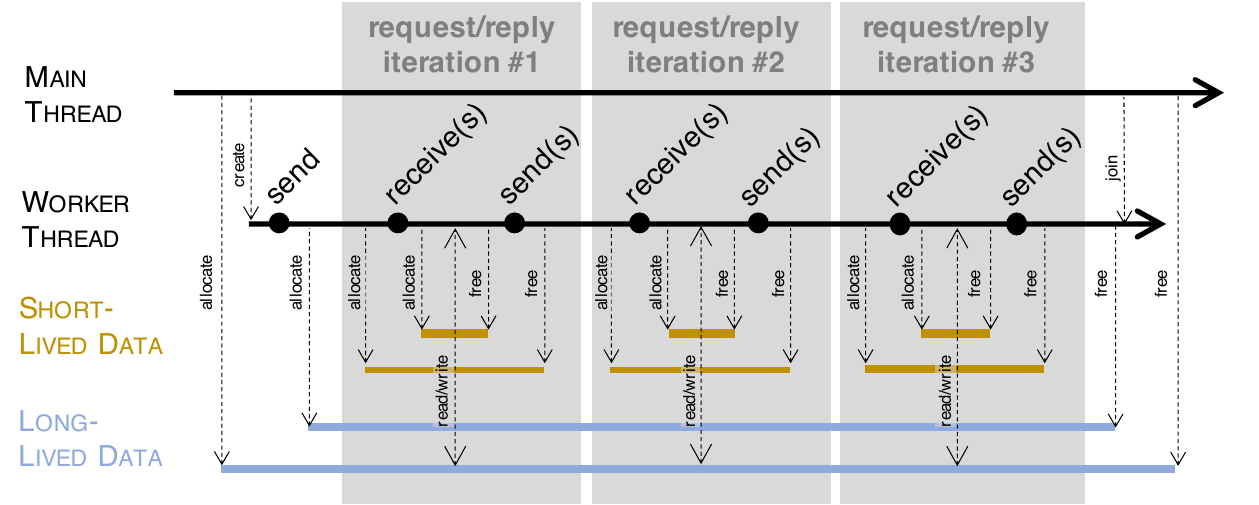}
  %\vspace{-0.2cm}
  \caption{The fundamental loop of network servers.}
  \label{fig:fundamental_loop}
  \end{center}
  %\vspace{-0.5cm}
\end{figure*}

\figurename{}~\ref{fig:fundamental_loop} shows the fundamental loop, with an overview of long- and short-lived data over a session. We refer to an individual request-reply exchange as an \emph{iteration} of the fundamental loop. For the purpose of example, in addition to the loop in the previous pseudocode, the figure also shows the typical case of a \emph{main} thread that listens for connection requests, and spawns a \emph{worker} thread for each session. Long-lived data can be allocated both by the main and the worker thread. At the beginning of a session, the worker may optionally perform a \textsc{send()} to transmit an initial \emph{banner} message that welcomes the client. Then, the worker performs one or more \textsc{receive()}s to get a request from the client, processes the request, and performs one or more \textsc{send()}s to communicate a reply to the client. The worker can allocate short-lived data both before the \textsc{receive()}s (e.g., a buffer for the incoming request) and after them (e.g., data for intermediate computations). Similarly, it can free short-lived data both before and after the \textsc{send()}s. We note that the end of a request/reply iteration (and the beginning of the next one) is denoted by a \textsc{receive()} after one or more \textsc{send()}s.

\stateafl{} has been designed on the basis of the fundamental loop of network servers. Similarly to \afl{} and other coverage-driven fuzzers, \stateafl{} is a mutation-based fuzzer, which automatically produces fuzz inputs by mutating previous ones, and gets feedback from the target program about the coverage achieved by the previous fuzz inputs. This feedback is important for coverage-driven fuzzers to prioritize which previous inputs to mutate, where to mutate them, and which mutation operators to apply. Differently from other fuzzers, \stateafl{} gets feedback not only about code coverage (e.g., which statements and branches were executed), but also about protocol states reached during an execution.

\begin{figure*}[!ht]
  \begin{center}
  \includegraphics[width=\columnwidth]{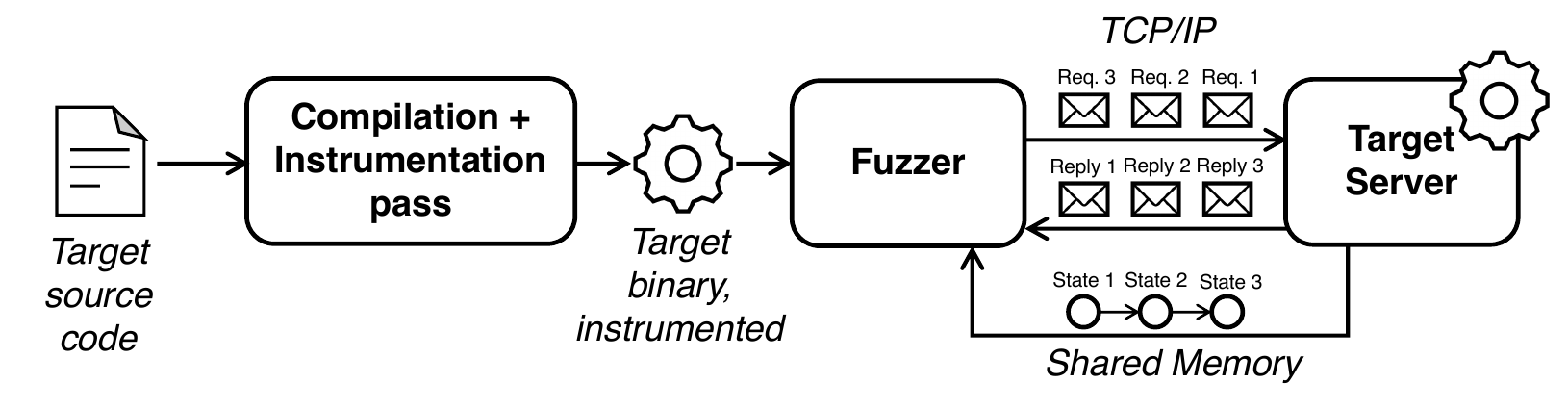}
  %\vspace{-0.4cm}
  \caption{Overview of \stateafl{}.}
  \label{fig:overview_stateafl}
  \end{center}
  %\vspace{-0.5cm}
\end{figure*}

\figurename{}~\ref{fig:overview_stateafl} provides an overview of \stateafl{}. In the first step, \stateafl{} compiles the source code of the target program. During this process, we apply \emph{compile-time instrumentation} techniques to introduce additional code in the binary executable generated by the compiler. The instrumentation adds code to collect feedback about the coverage of protocol states, in a similar manner to instrumentation code added by other fuzzers for analyzing code coverage. 
Note that this approach requires the availability of the source code of the target server. This scenario represents relevant use cases for stateful network fuzzing, such as developers that need to fuzz their own software (e.g., as part of an automated V\&V process), and users of open-source software that need to gain additional security evidence. We leave the fuzzing of binary-only software out of the scope of this work.

After the instrumentation, the \stateafl{} fuzzer runs the target server by launching the binary. To exercise the target server with fuzz inputs, \stateafl{} exchanges TCP/IP messages with the target, in the same way of a client. A fuzz input is managed as a sequence of request messages: for each request message in the sequence, the fuzzer sends it through TCP/IP, waits for a reply message, and moves to the next request message. 
In addition, the \stateafl{} fuzzer collects information about protocol states reached by the target server, through a side channel (a shared memory area). The feedback consists of a sequence of states, one for each request/reply iteration. 
Note that StateAFL works on application-level messages, as demarked by the fundamental loop of \textsc{send()} and \textsc{receive()} primitives. The application messages may be (or may be not) divided among multiple packets by the TCP/IP stack, transparently to the fuzzer.

Our current design focuses on TCP/IP (including both the TCP and UDP transport protocols), since this protocol suite is the most commonly adopted by network servers. It is possible to easily adapt the design to other communication protocols, such as RPC servers. This design also focuses on client-server communication; fuzzing through multiple channels (e.g., multi-party protocols) represents a separate, still open research problem \citep{profuzzbench2021}, which we leave out of scope of this paper.

\subsection{Instrumentation probes}

To collect feedback about protocol states, compile-time instrumentation weaves probes into the code of the target server. Probes are inserted at specific points of the code that allocate and free memory, and that send and receive data on the network. A probe consists of a call instruction, which invokes an external function, in order to perform actions when the server executes the instrumented points of interest. In some cases, the probe passes run-time information about the process to the external function (e.g., the address and size of a memory area).

The compile-time instrumentation links the target program to a library provided by \stateafl{}, which contains the external functions to be invoked by the probes. These library functions will collect and analyze data for inferring protocol states. In particular, the library provides the following functions:

\begin{itemize}

\item \textsc{on\_allocate}: This function is invoked when a heap or stack memory area has been allocated (e.g., using \textsc{malloc}). It takes in input the address and size of the memory area. It keeps track of all data structures, regardless that they are long- or short-lived (which cannot be determined at the moment of the allocation, but only afterwards).

\item \textsc{on\_free}: This function is invoked when a heap or stack memory area has been deallocated (e.g., using \textsc{free}). It takes in input the address of the memory area. The function updates the status of data structures that were tracked by \textsc{on\_allocate}.

\item \textsc{on\_send} and \textsc{on\_receive}: These functions are invoked when the server transmits or receives data to/from the client (e.g., a write or read on a socket), and keep track of the fundamental loop of the network server.

\item \textsc{on\_process\_start}: Executes at the start-up of the network server. It initializes the internal data structures (e.g., \textsc{alloc\_records\_map} and \textsc{alloc\_dumps\_queue}), the internal state machine, and the shared memory area to communicate with the fuzzer.

\item \textsc{on\_process\_end}: Executes at the termination of the network server. It analyzes the data structures that were allocated by the network server during its execution, identifies which data are long-lived, and computes the sequence of protocol states, to be shared with the fuzzer.

\end{itemize}

\begin{figure}
  \begin{center}
  \includegraphics[width=0.5\columnwidth]{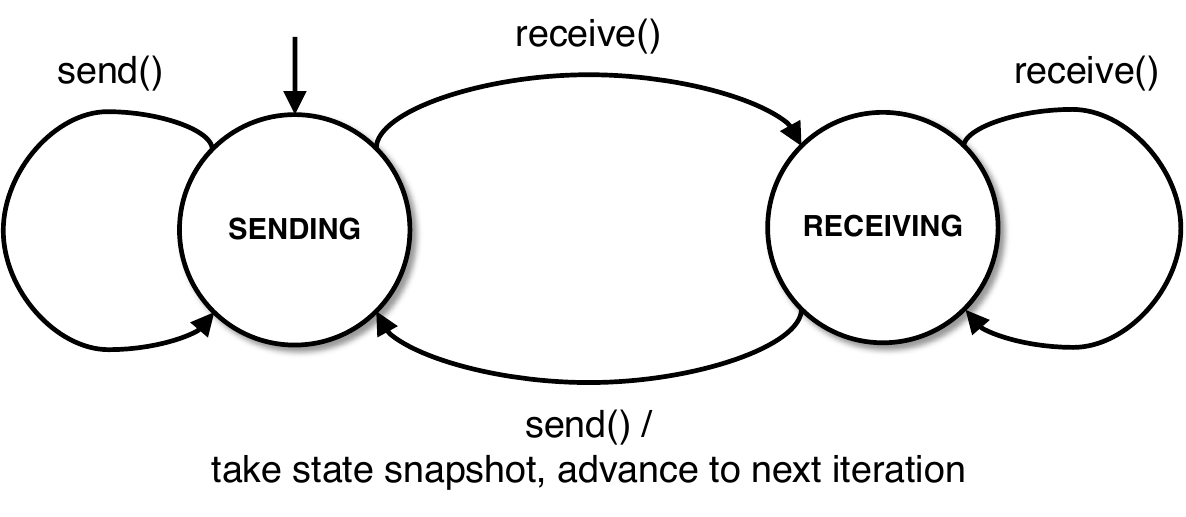}
  %\vspace{-0.6cm}
  \caption{State machine to keep track of protocol iterations.}
  \label{fig:state_machine}
  %\vspace{-0.7cm}
  \end{center}
\end{figure}

\stateafl{} keeps track of the iterations of request/reply exchanges, by probing send()s and receive()s made by the network server. On these operations, \textsc{on\_receive} and \textsc{on\_send} update an internal state machine according to \figurename{}~\ref{fig:state_machine}. These functions (Algs.~\ref{alg:on_receive} and \ref{alg:on_send}) represent the current iteration using the global integer variable $current\_iter\_no$, allocated by the \stateafl{} library and initially set to 0 (Alg.~\ref{alg:on_process_start}). The state machine identifies the end of an iteration, by looking for a series of receive()s and, after some processing of the request, a send() (or, the first of a sequence of send()s). By the time that the server starts to send a reply, the long-lived data have been updated by the network server, and reflect a new protocol state. Therefore, on the first send() event, the iteration is considered as terminated, and a new one as started. We update the current iteration, by increasing $current\_iter\_no$ by one. Please note that the lifetime of short-lived data could end right before or right after the end of an iteration, depending on the network server. However, this does not pose a problem for \stateafl{}, since short-lived data are going to be ignored by later analysis, regardless of which iterations they span over.

\begin{algorithm}[!htb]
\caption{Triggered when the server starts.}
\label{alg:on_process_start}
\begin{algorithmic}[1]

\Procedure{on\_process\_start}{}

\State $alloc\_records\_map \gets $ new map
\State $alloc\_dumps\_queue \gets $ new queue
\State $ current\_iter\_no \gets 0$

\EndProcedure

\end{algorithmic}
\end{algorithm}

\begin{algorithm}[!htb]
\caption{Triggered when the server retrieves a message.}
\label{alg:on_receive}
\begin{algorithmic}[1]

\Procedure{on\_receive}{}
\If { state machine is $SENDING$ }

    \State state machine $\gets RECEIVING$

\EndIf
\EndProcedure

\end{algorithmic}
\end{algorithm}

\begin{algorithm}[!htb]
\caption{Triggered when the server sends a message.}
\label{alg:on_send}
\begin{algorithmic}[1]

\Procedure{on\_send}{}

\If { state machine is $RECEIVING$ }

    \State state machine $\gets SENDING$ 
    
    \State $dump\_current\_state()$
    
    \State $current\_iter\_no \gets current\_iter\_no + 1$

\EndIf

\EndProcedure

\end{algorithmic}
\end{algorithm}

During the execution, when a memory area is allocated on the heap or on the stack, the probes trigger the function \textsc{on\_allocate} (Alg.~\ref{alg:on_allocate}). This function records the allocation using the \textsc{alloc\_record} data structure, which includes: (i) the number of the iteration at which the memory area was allocated, (ii) the number of the iteration at which it was deallocated (to be filled by \textsc{on\_free}), (iii) the address of the memory area, and (iv) the size of the memory area. The \textsc{alloc\_record} data structure is stored into a map (i.e., an associative array), using as key the address of the memory area.
The memory area is initialized to zero: \stateafl{} relies on the contents of the memory area as a proxy for the current protocol state, and must not contain random data. Since heap and stack memory areas in standard C are not automatically initialized, their contents are unpredictable and not correlated to the protocol state, until they are written by the program. Therefore, we initialize the heap and stack memory areas to assure that their unused parts have still a fixed and predictable value, which does not mislead the inference of protocol states. 
When the memory area is freed, the \textsc{on\_free} function updates its \textsc{alloc\_record} structure with the iteration number at which the area was freed (Alg.~\ref{alg:on_free}). Still, the \textsc{alloc\_record} structure lasts until the termination of the network server. As an optimization, Alg.~\ref{alg:on_allocate} only records allocations that are made during the first iteration. The allocations made by the subsequent iterations do not span the entire lifetime of the process, and are considered as short-lived.

\begin{algorithm}[!htb]
\caption{Triggered when the server allocates memory.}
\label{alg:on_allocate}
\begin{algorithmic}[1]

\Procedure{on\_allocate}{address, size}

\If { $current\_iter\_no = 0$ }

\State {$a \gets$ new $alloc\_record$}

%\item[]
\vspace{0.2cm}
    
\State $a.iter\_no\_init \gets current\_iter\_no$
\State $a.iter\_no\_end \gets -1$
\State $a.addr \gets $ initial address of allocated area
\State $a.size \gets $ size of allocated area

%\item[]
\vspace{0.2cm}

\State $alloc\_records\_map.put( a.addr, a )$

\State zero-initialize the allocated area

\EndIf

\EndProcedure

\end{algorithmic}
\end{algorithm}

\begin{algorithm}[!htb]
\caption{Triggered when the server frees memory.}
\label{alg:on_free}
\begin{algorithmic}[1]

\Procedure{on\_free}{address}

\State $a \gets alloc\_records\_map.get( address )$

\State $a.iter\_no\_end \gets current\_iter\_no$

\State $alloc\_records\_map.remove( a.addr )$

\EndProcedure

\end{algorithmic}
\end{algorithm}

The \textsc{alloc\_record} data structures are inspected by \stateafl{} when the current iteration terminates, and the state machine moves to the next iteration (i.e., the transition from \textsc{RECEIVING} to \textsc{SENDING}, see \figurename{}~\ref{fig:state_machine}). On this event, the \textsc{on\_send} function calls \textsc{dump\_current\_state} (Alg.~\ref{alg:dump_current_state}), which iterates over  all of the currently-allocated heap and stack areas in \textsc{alloc\_records\_map}.

The \textsc{dump\_current\_state} function takes a snapshot (a \emph{dump}) of the contents of every memory area, by saving them into an \textsc{alloc\_dump} data structure. Moreover, the \textsc{alloc\_dump} will track the iteration number at which the snapshot was taken, and a reference to the \textsc{alloc\_records\_map} for the memory area. Even if the network server deallocates the memory area, its \textsc{alloc\_record} structure is still saved and referenced by the \textsc{alloc\_dump} structure. All \textsc{alloc\_dump} structures are enqueued into \textsc{alloc\_dumps\_queue}.

\begin{figure*}[!t]
  \begin{center}
  \includegraphics[width=\columnwidth]{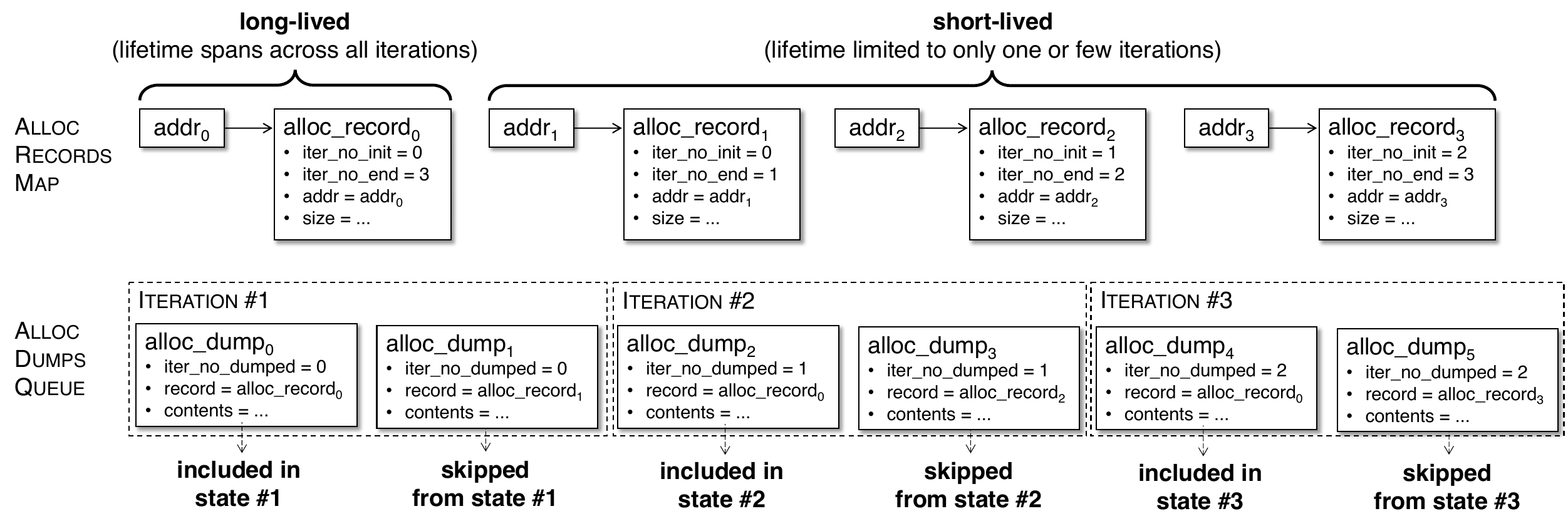}
  \caption{Example of data structures ($alloc\_record$ and $alloc\_dump$), after an execution with three iterations, with one long-lived area allocated at the beginning and freed at the end, and three short-lived areas allocated and freed at each iteration.}
  \label{fig:example_data_structures}
  \end{center}
  %\vspace{-0.5cm}
\end{figure*}

\begin{algorithm}[!htb]
\caption{Takes snapshots of memory areas}
\label{alg:dump_current_state}
\begin{algorithmic}[1]

\Procedure{dump\_current\_state}{}

\ForAll {$a \in alloc\_records\_map$}
    \State $d \gets$ new $alloc\_dump$
    \State $d.iter\_no\_dumped \gets current\_iter\_no$
    \State $d.record \gets $ reference to $a$
    \State $d.contents \gets $ copy $a.addr$'s contents
    
    %\item[]
    \vspace{0.2cm}
    
    \State $alloc\_dumps\_queue.push(d)$
\EndFor

\EndProcedure

\end{algorithmic}
\end{algorithm}

\figurename{}~\ref{fig:example_data_structures} provides an example of the \textsc{alloc\_record} and \textsc{alloc\_dump} data structures. In this example, the network server initially allocates a long-lived data structure at address $addr_0$, and represented by $addr\_record_0$. For this long-lived area, $iter\_no\_init$ is initialized to 0, since it has been allocated before the first iteration could complete. Then, the network server iterates for three request/reply exchanges. At every iteration, the server allocates a short-lived data structure before processing the request, and deallocates it after sending the reply. Thus, the server allocates in total 3 short-lived memory areas ($addr\_record_1$, $addr\_record_2$, and $addr\_record_3$, respectively). The alloc\_dump structures for the short-lived data are annotated with the iteration in which they were allocated ($iter\_no\_init = 0, 1, 2$, respectively) and deallocated ($iter\_no\_end = 1, 2, 3$, respectively). We remark that allocations performed after the first iteration ($1, 2, \ldots$) are not actually tracked by our algorithm, but are included in this discussion as an example of short-lived data.

In the example of \figurename{}~\ref{fig:example_data_structures}, the \textsc{dump\_current\_state} function is triggered 3 times, at the end of each iteration. At the first iteration, \textsc{dump\_current\_state} dumps the current contents of the long-lived data structure ($alloc\_dump_0$), and the contents of the first short-lived data structure ($alloc\_dump_1$). Note that the \textsc{alloc\_dump} structures have a reference to the \textsc{alloc\_record} structures. Similarly, at the end of the second and third iterations, \textsc{dump\_current\_state} dumps again the current contents of the long-lived data structure ($alloc\_dump_2$ and $alloc\_dump_4$). The dumps $alloc\_dump_0$, $alloc\_dump_2$, and $alloc\_dump_4$ are from the same long-lived data structure, but they can hold different contents, as the network server updates long-lived data at each iteration. Finally, the \figurename{}~\ref{fig:example_data_structures} includes the dumps $alloc\_dump_3$ and $alloc\_dump_5$ for the other two short-lived area, taken respectively at the end of the second and third iteration.

\subsection{Post-execution analysis}
\label{subsec:post-exec_analysis}

The dumps in \textsc{alloc\_dumps\_queue} are later analyzed at the end of the execution, after that all request/reply iterations for the fuzz input have been completed. After the last iteration, the network server is forcefully terminated, and the \textsc{on\_process\_end function} is triggered (Alg.~\ref{alg:on_process_end}). In turn, it calls the \textsc{save\_state\_seq function}.

\begin{algorithm}[!htb]
\caption{Triggered when the server terminates.}
\label{alg:on_process_end}
\begin{algorithmic}[1]

\Procedure{on\_process\_end}{}

\State $total\_iterations \gets current\_iter\_no$

\State $save\_state\_seq()$

\EndProcedure

\end{algorithmic}
\end{algorithm}

The \textsc{save\_state\_seq} function (Alg.~\ref{alg:save_state_seq}) iterates over the \textsc{alloc\_dumps\_queue}. As result, \textsc{save\_state\_seq} generates a sequence of states, with one state for each iteration made by the network server. A state is represented by a unique integer value (\emph{state id}), based on the contents of long-live data at the end of the iteration. Therefore, if long-lived data are updated between an iteration and the next one, the two states will be represented by two distinct integer values. Otherwise, if the long-lived data stay unchanged between iterations, the states are represented by the same integer value. Of course, it is possible that the same integer value (i.e., the same state) appears multiple times at distant times in the sequence, as the network server can return to a previous state, depending on the server behavior. In the example of \figurename{}~\ref{fig:example_data_structures}, \textsc{save\_state\_seq} generates a sequence of three states, represented by three integer values, which can be different or identical depending on any changes made to the long-lived data structure.

%For every iteration, the algorithm analyzes the union of all dumps of long-lived data in that iteration. The algorithm aims to assign a \emph{state id} that is uniquely mapped to the contents of the dumps. For example, if the memory contents at the current iteration are identical to the contents of the previous iteration, then the algorithm should return the same \emph{state id} for both iterations. Similarly, if the target server goes back to the same memory contents of a previous iteration, the \emph{state id} should be the same of that iteration. Instead, if the current iteration changes the memory contents (e.g., the last message affects the current protocol state), then the algorithm should return a new, unique \emph{state id} for the iteration.

\begin{algorithm}[!htb]
\caption{Generates a sequence of protocol states.}
\label{alg:save_state_seq}
\begin{algorithmic}[1]

\Procedure{save\_state\_seq}{}

\State $states\_sequence \gets $ new list of integers

\State $prev\_iter\_no \gets 0$
\State $tlsh\_hash \gets 0$

%\item[]
\vspace{0.2cm}
    
\ForAll {$d \in alloc\_dumps\_queue$}

    %\item[]

    \vspace{0.2cm}

    \State \(\triangleright\) Long-lived data span across all iterations
    \If {$d.record.iter\_no\_init > 0$ or \\
        \quad \quad \quad \quad ($d.record.iter\_no\_end < total\_iterations$ and \\ \quad \quad \quad \quad $d.record.iter\_no\_end \ne -1$)}
        \State skip $d$     \Comment{Ignore short-lived data}
    \EndIf
    
    %\item[]
    \vspace{0.2cm}
    
    \If {$d.iter\_no\_dumped > prev\_iter\_no$}

       \State \(\triangleright\) Save state of the current iteration before the next

        \State $state\_id \gets get\_state\_id(tlsh\_hash)$
    
        \State $states\_sequence.push(state\_id)$
        
        \State $tlsh\_hash \gets 0$
        
    \EndIf
    
    %\item[]
    \vspace{0.2cm}

    \State $update\_tlsh(tlsh\_hash, d.contents)$
    \State $prev\_iter\_no \gets d.iter\_no\_dumped$
    
\EndFor

%\item[]
\vspace{0.2cm}

\State $state\_id \gets get\_state\_id(tlsh\_hash)$
\State $states\_sequence.push(state\_id)$
\State $save\_to\_shared\_memory(states\_sequence)$

\EndProcedure

\item[]

\Procedure{get\_state\_id}{$tlsh\_hash$}

\State $radius \gets \epsilon$

\State $state\_id \gets mvptree\_lookup(tlsh\_hash, radius)$

%\item[]
\vspace{0.2cm}
    
\If { $state\_id = \varnothing$ } 

	\State $state\_id \gets mvptree\_count() + 1$

	\State $mvptree\_add( \{ tlsh\_hash, state\_id \} )$
\EndIf

%\item[]
\vspace{0.2cm}
    
\State \Return $state\_id$

\EndProcedure

\end{algorithmic}
\end{algorithm}

Alg.~\ref{alg:save_state_seq} iterates over \textsc{alloc\_dump}s. The algorithm identifies dumps of long-lived data, by looking for those whose lifetime spans across all iterations. The algorithm skips a dump as short-lived data if its memory area has been allocated after the first iteration ($iter\_no\_init > 0$), or if it has been deallocated before the termination of the last iteration ($iter\_no\_end < total\_iterations$, except $iter\_no\_end = -1$ that denotes an area never deallocated). In the case of \figurename{}~\ref{fig:example_data_structures}, the first state is obtained only from $alloc\_dump_0$ (i.e., the first dump of the long-lived area); similarly, the second and third states are only based on $alloc\_dump_2$ and $alloc\_dump_4$ (i.e., second and third dump of the long-lived area). 

When iterating over the dumps, the algorithm computes a \emph{hash function} over the union of all dumps for the same protocol iteration. The hash value is adopted to map the memory contents to a unique state identifier. The hash value is computed incrementally, by updating it with one dump at a time. 
When the algorithm finds a dump for a new iteration ($d.iter\_no\_dumped > prev\_iter\_no$), the state identifier for the previous iteration is finalized and pushed to the sequence, and the analysis is repeated for the next iteration, until all dumps have been analyzed. 

A potential drawback of using hash functions is that the state identifier could be over-sensitive to small, negligible variations of the memory contents not correlated with the protocol state, because of non-deterministic factors. For example, as the fuzzer executes the target server multiple times, the process may get from the OS different descriptors for socket and file I/O, or its data may be allocated at different addresses of the virtual memory space. In turn, these values can be copied to long-lived data structures (e.g., pointer variables). Most hash functions are designed to be sensitive to small changes, and to generate largely different hash values even if the inputs are similar. Therefore, small, non-deterministic variations would lead to different, redundant state identifiers, even if the variations do not affect the behavior of the server. Disabling OS randomization mechanisms reduces, but does not prevent such variations.

To mitigate this issue, our algorithm adopts \emph{Locality-Sensitive Hashing} (LSH). In LSH, two similar inputs (e.g., differing only for few bits) result in two hash values that are different, but similar \citep{jafari2021survey}. This form of hashing has applications in several domains, such as document retrieval, plagiarism detection, and bioinformatics. In the field of software security, LSH has most often been adopted for analyzing malware similarity \citep{oliver2013tlsh,ali2020scalable}. In one previous work, LSH has been used on path constraints of symbolic execution states, in order to speed-up the search for previously-solved states \citep{cady2017tree}. 
In general, LSH enables the quick look-up of items that are similar to the one under analysis, by looking for items with a similar hash value according to some distance metric.

In particular, in this work we adopt the \emph{Trend Micro Locality Sensitive Hash} (TLSH), a popular algorithm that has shown high robustness against small differences in the inputs \citep{oliver2013tlsh,ali2020scalable}. TLSH computes a distribution of the bit patterns in the data, and generates a digest from this distribution. TLSH also comes with a distance metric between hash values, which approximates the Hamming distance between two hash digest bodies. 
The function \textsc{get\_state\_id} (Alg.~\ref{alg:save_state_seq}) takes in input the TLSH hash of long-lived data for the current iteration, and performs a nearest neighbor search for the previous most similar hash value, within a maximum distance of $\epsilon$. If a similar hash value is found, the algorithm returns its mapped state identifier; otherwise, a new pair $\{tlsh\_hash, state\_id\}$ is stored using a new, unique state identifier. The algorithm uses a \emph{Multi-Vantage Point} (MVP) tree data structure to store the pairs, and to perform look-ups based on the TLSH hash and on the TLSH distance metric. We use a MVP tree for computationally-efficient nearest-neighbor search, as it avoids expensive pair-wise comparisons with previous values in the tree \citep{bozkaya1997distance,bozkaya1999indexing}.

The distance threshold $\epsilon$ is dynamically calibrated for the target server under test, according to Alg.~\ref{alg:calibration}. Before fuzzing, \stateafl{} performs a calibration stage, by executing the server multiple times using the seed inputs, and by computing a sequence of hash values at each iteration of each repetition. In our implementation, we run one ``reference'' execution plus 3 additional repetitions for each seed input. Since the server is executed with the same inputs, the distance threshold $\epsilon$ should be calibrated such that the sequence of state identifiers is the same across repetitions. Therefore, the calibration stage compares the hash value at each iteration of the first execution ($reference\_hashes\_seq[i]$) with the corresponding hash value of every other repetition ($new\_hashes\_seq[i]$), and collects the distances between the hash values ($distances$). 
Then, it takes the $90^{th}$ percentile of these distances as a conservative choice for $\epsilon$. The rationale for choosing the $90^{th}$ percentile is that the distances across calibration runs should fall within the threshold (i.e., clustered in the same state), since these runs process the same requests starting from the same initial state. Taking a higher threshold (e.g., the maximum distance) would be too conservative, since it would be influenced by sporadic outliers caused by non-determinism. Taking a lower threshold (e.g., a lower percentile) would make the approach prone to redundant states, since two occurrences of the same state may not be recognized as such.

After calibrating $\epsilon$, it is expected that the server often returns to a previously known state, and that it visits new states infrequently (e.g., the fuzzer triggers a new corner case in the protocol). 
However, it is possible that the threshold is set too low if the initial seed inputs are too few or too short, since only few states are visited during the calibration. This case often occurs when a security assessor undertakes a fuzzing campaign for a new or unfamiliar system. Therefore, \stateafl{} provides a threshold adjustment procedure as a fallback option to compensate for an under-estimated threshold. The adjustment is based on the observation that, if the threshold is too low, then a high number of new states will quickly occur during fuzzing. 
To handle this case, the fuzzer increases the threshold when new states are added for 5 consecutive fuzz inputs, which is very unlikely to happen for a well-calibrated threshold, since new states should be infrequent. The threshold is increased by 10, which is a relatively low value to account for small corrections, but also not too low, in order to react quickly. 
As in the original proposal of TLSH, we vary the threshold $\epsilon$ between $5$ and $100$.

\begin{algorithm}[!htb]
\caption{Threshold calibration.}
\label{alg:calibration}
\begin{algorithmic}[1]

\Procedure{threshold\_calibration}{}

\State $distances \gets $ empty set

\vspace{0.2cm}
    
\ForAll {$s \in seed inputs$}

    %\item[]

    \vspace{0.2cm}
    
    \State $run\_target(s)$
    
    \State $reference\_hashes\_seq \gets get\_last\_run\_hashes\_seq()$
    
    \vspace{0.2cm}

    \ForAll {$r \in 1 \ldots \# \textrm{repetitions}$}

        \vspace{0.2cm}

        \State $run\_target(s)$
        
        \State $new\_hashes\_seq \gets get\_last\_run\_hashes\_seq()$
        
        \vspace{0.2cm}

        \ForAll {$i \in 1 \ldots len(reference\_hashes\_seq)$}
              \State $dist \gets tlsh\_distance(reference\_hashes\_seq[i], ~ new\_hashes\_seq[i])$

              \State $distances \gets distances \cup dist$
       
        \EndFor

        \vspace{0.2cm}
        
    \EndFor

    \vspace{0.2cm}

\EndFor

\vspace{0.2cm}

\State  \(\triangleright\) Get $90^{th}$ percentile of distances
\State $threshold \gets distances_{90\%}$

\vspace{0.2cm}

\EndProcedure

\end{algorithmic}
\end{algorithm}

Ultimately, the \textsc{save\_state\_seq} function returns the sequence of states to the \stateafl{} fuzzer. The fuzzer incrementally grows a state machine after each fuzz input, based on the returned sequence of states. For example, when a fuzz input covers a new state, the state machine is updated by adding a new state and a new transition from the previous state in the sequence. Similarly, a new transition is added to the state machine when a pair of states appears consecutively in a sequence for the first time.

\begin{figure}[!ht]
  \begin{center}
  \subfloat[Inferred protocol state machine.\label{fig:example-state-machine-fsm}]{%
        \raisebox{0mm}{\includegraphics[width=0.6\columnwidth]{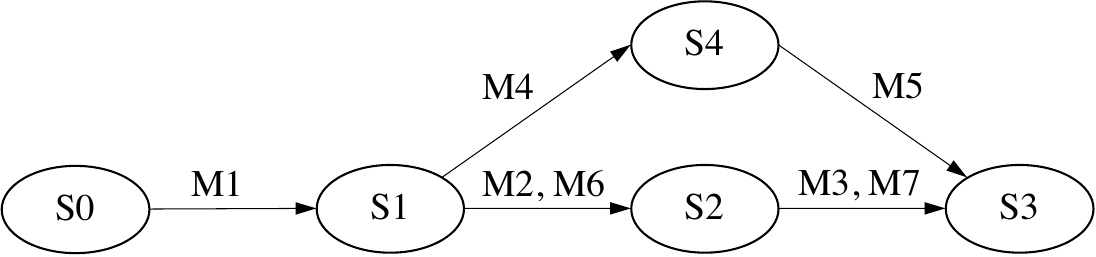}}%
  }
  \qquad
  \subfloat[Information associates to the states.\label{fig:example-state-machine-table}]{%
        \includegraphics[width=\columnwidth]{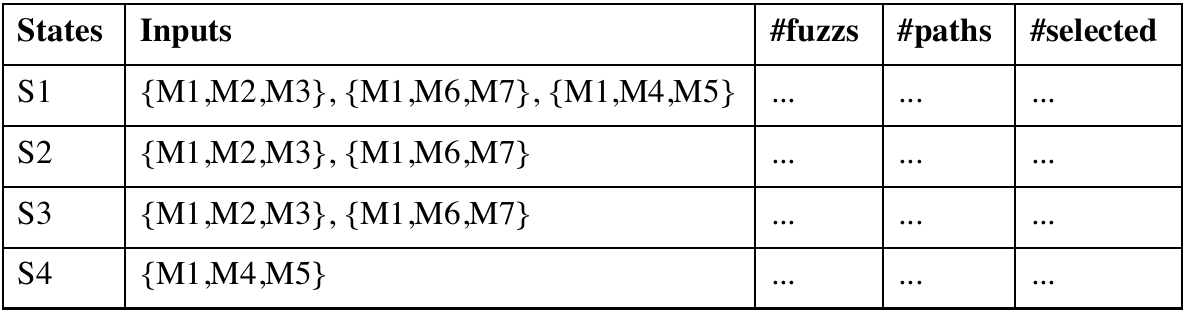}%
  }
  \end{center}
  \caption{Example of protocol state machine inference and state selection.}
  \label{fig:example-state-machine-and-selection}
\end{figure}

During the fuzzing process, \stateafl{} uses the state machine to generate new fuzz inputs, in order to further increase code coverage and to explore the protocol. Heuristics from previous model-based fuzzing techniques can be leveraged for this purpose \citep{pham2020aflnet}, to: (i) select a target state from the state machine; (ii) identify a previous fuzz input that reached the selected state; and (iii) apply a mutation operator on the message that is sent from the target state. 
%(rarer ones are selected with higher probability)
%according to how often states have been exercised, and how often they contributed to discover new states or to cover new paths.
\figurename{}~\ref{fig:example-state-machine-and-selection} provides an example of inferred protocol state machine, and related information that is tracked by the fuzzer for state selection purposes: \emph{\#fuzzs} is the number of previous mutated inputs that have exercised that state; \emph{\#paths} is the number of times that the code or state coverage increased when the state was previously selected; and \emph{\#selected} is the number of times that the state was previously selected. Moreover, the table tracks inputs that covered each state and that are ``interesting'', i.e., that increased code or state coverage. The target state is selected with a probability that is inversely proportional to \emph{\#fuzzs} and \emph{\#selected}, and proportional to \emph{\#paths}. After selecting a state, the fuzzer randomly selects one of the previous interesting inputs that covered that state (\emph{Inputs}). Finally, the fuzzer identifies the message in the input that reaches the selected state, and it targets for mutation the subsequent message in the same input. For example, in \figurename{}~\ref{fig:example-state-machine-and-selection}, if the fuzzer targets \emph{S4}, it will generate a fuzz input beginning with messages \emph{M1} and \emph{M4}, followed by a mutated version of message \emph{M5}.

\begin{figure}[!ht]
  \begin{center}
  \includegraphics[width=\columnwidth]{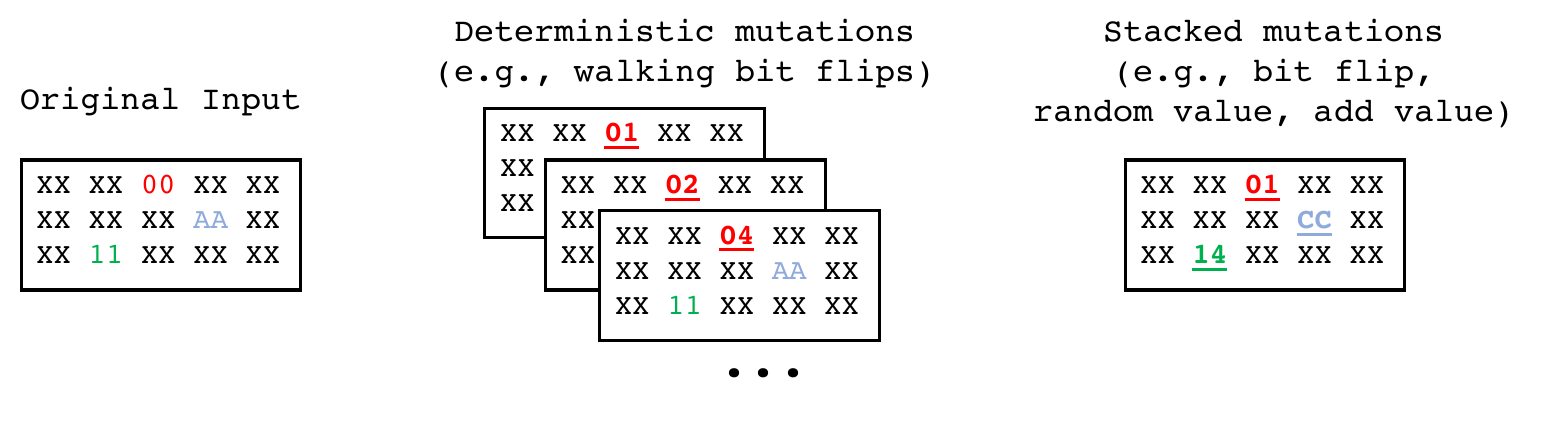}
  %\vspace{-0.2cm}
  \caption{Examples of mutated inputs.}
  \label{fig:mutation-examples}
  \end{center}
\end{figure}

The mutations include both byte-level operators and message-level ones (\tablename{}~\ref{tab:mutation-operators}). The byte-level operators are derived from the \afl{} fuzzer \citep{afl}, and modify the content of an individual message. 
There are two types of byte-level mutation operators: \emph{deterministic} and \emph{stacked}. The deterministic ones systematically mutate all of the bits, bytes, words, and double words in the original input. For example, in the ``single walking bit flip'' mode, the fuzzer iterates sequentially over all of the bits in the original input, and generates a distinct fuzz input by inverting each of these bits. \figurename{}~\ref{fig:mutation-examples} shows the case of three different fuzz inputs generated by inverting three different bits of the original input. This approach is meant to discover sensitive parts of the original input (e.g., headers) that increase coverage when mutated, and that are interesting to further mutate. Afterwards, the fuzzer applies \emph{stacked} mutations, where multiple types of mutations are applied on the same fuzz input. In this case, the mutated parts of the inputs are randomly selected. For example, in \figurename{}~\ref{fig:mutation-examples} three mutation operators are applied on the same input, respectively a random bit flip, a random byte set to a random value (\texttt{CC}), and a random value added to a random byte (\texttt{+3}).  
Four more mutation operators at the message-level are derived from the \aflnet{} fuzzer \citep{pham2020aflnet}. These mutations replace, insert, and duplicate a message at the location of the message for selected fuzzing. The message-level operators are stacked with byte-level operators.

\begin{table}[] 
\centering
\caption{Mutation operators.}
\label{tab:mutation-operators}
%\vspace{-0.2cm}
{
%\small
\footnotesize
\begin{tabular}{lll}
\toprule
\textbf{Mutation} & \textbf{Type} & \textbf{Fuzzer} \\
\midrule
Single, two, or four walking bit flips & Deterministic & \afl{} \\ 
One, two, or four walking byte flips & Deterministic & \afl{} \\ 
Walking 8-, 16-, or 32-bit arithmetics & Deterministic & \afl{} \\ 
Walking 8-, 16-, or 32-bit innteresting values & Deterministic & \afl{} \\ 
Walking overwrite with user-supplied dictionary values & Deterministic & \afl{} \\ 
Walking insertion of user-supplied dictionary values & Deterministic & \afl{} \\ 
Splicing multiple inputs & Deterministic & \afl{} \\ 
Flip single random bit & Stacked & \afl{} \\ 
Set random byte, word, or double word to interesting value & Stacked & \afl{} \\ 
Subtract value at a random byte, word, or double word & Stacked & \afl{} \\ 
Add value at a random byte, word, or double word & Stacked & \afl{} \\ 
Set random byte to random value & Stacked & \afl{} \\ 
Delete random bytes & Stacked & \afl{} \\ 
Clone random bytes & Stacked & \afl{} \\ 
Insert block of constant bytes in random position & Stacked & \afl{} \\ 
Overwrite bytes with randomly selected ones & Stacked & \afl{} \\ 
Overwrite bytes with fixed bytes & Stacked & \afl{} \\ 
Replace message with a random one from a random input & Stacked & \aflnet{} \\ 
Insert random message from a random input, before the target message & Stacked & \aflnet{} \\ 
Insert random message from a random input, after the target message & Stacked & \aflnet{} \\ 
Duplicate message & Stacked & \aflnet{} \\ 
\bottomrule
\end{tabular} 
}
\end{table}

\subsection{Implementation}
\label{subsec:implementation}

We implemented \stateafl{} on top of the codebase of \afl{} and \aflnet{}. For compile-time instrumentation, we extended the \textsc{afl-clang-fast} utility, which is provided by \afl{} to compile the target program, and which adds a compiler pass to introduce instructions for coverage profiling. In the compiler pass, we add further instrumentation to introduce probes in the program, as discussed in Section~\ref{sec:proposed}. 
We focus on the case where the source code of the target server is available for fuzzing.

Probes are injected on heap allocation sites that invoke the standard C library functions \textsc{malloc}, \textsc{realloc}, \textsc{calloc} and \textsc{free}, and the C++ operators \textsc{new} and \textsc{delete}, in order to call \textsc{on\_allocate} and \textsc{on\_free}. The probes take the size of the allocated memory area from the input of the allocation, and its memory address from the output. Similarly, allocation sites of stack memory are probed, by identifying \textsc{push} and \textsc{pop} operations on the stack that modify the stack pointer register. The probes compute the address and size of the allocated memory area from the stack pointer register. In order to avoid excessive overhead that may be caused by probing all stack allocations, we only probe allocations of data structures larger than a threshold (64 bytes), since in practice small allocations typically represent temporary variables and do not hold long-lived data structures. Data in globals and TLS are handled as memory areas with a lifetime spanning the entire execution.

Probes are also injected on call sites to standard library functions that send and receive network data, such as \textsc{send}, \textsc{sendto}, and \textsc{sendmsg} (to trigger \textsc{on\_send}), \textsc{recv}, \textsc{recvfrom}, and \textsc{recvmsg} (to trigger \textsc{on\_receive}). We also probe the standard library functions \textsc{read}, \textsc{write}, \textsc{fprintf}, \textsc{fgets}, \textsc{fread}, and \textsc{fwrite}, with an additional check that the file descriptor in input is a network socket. We allow the user to specify (using an environment variable) any program-specific function that should be instrumented for intercepting network communication. For example, for a server that implements a HTTP-based protocol using the \emph{libevent} API, the user can instruct \stateafl{} to instrument the \textsc{evhttp\_request\_*} and \textsc{evhttp\_send\_*} API functions, to trigger the external functions \textsc{on\_receive} and \textsc{on\_send}, respectively. Finally, our probes invoke \textsc{on\_process\_start} and \textsc{on\_process\_end} on start-up and termination of the target program.

After the compiler pass, we link the program executable with a library that implements the event handlers, to be called by the probes. The library shares a UNIX SysV shared memory to exchange state sequences. We replaced the protocol-specific message parsers of \aflnet{} with a single, generic function that read the state sequence from the shared memory, without parsing response codes from the messages. We reuse the test automation from \afl{} and \aflnet{} to execute the target program (e.g., the fork server), and to mutate fuzz inputs.

As an optimization for speeding-up fuzzing, \stateafl{} can be configured to perform heavy-weight post-execution analysis of long-lived memory only when strictly needed. Fuzz inputs are normally processed without performing the post-execution analysis, to have a high fuzzing throughtput; when a fuzz input covers a new program path (i.e., increases the code coverage), it is processed again in order to the post-execution analysis. The analysis returns a sequence of states reached by the fuzz input, which is stored by the fuzzer. This information is used by the fuzzer to generate more fuzz inputs starting from each state in the sequence.

\section{Experimental plan}
\label{sec:evaluation_plan}

We evaluate \stateafl{} by fuzzing real-world network servers from popular open-source projects. The experimental plan addresses the following research questions:

\vspace{3pt}
%\noindent
\textbf{RQ1: How \stateafl{} compares to state-of-the-art network fuzzing?} We evaluate them with respect to both code coverage, which is a typical indicator of the depth of fuzz testing, and crashes of the targets, which indicates that a fuzzer can uncover potential security issues \citep{klees2018evaluating}.

\vspace{3pt}
%\noindent
\textbf{RQ2: How accurate are the inferred protocol states?} This is a difficult question, since we lack a ground truth for the protocols to be inferred, and since the protocol state machine depends on the specific protocol implementation of the network server \citep{poll2015protocol}. We address this question through a qualitative analysis on one of the target network servers, by manually analyzing its source code, to check that the inferred states are not redundant and reflect the expected behavior of the protocol. As a further term of comparison, we also compare the inferred state machines by \stateafl{} with the ones inferred by custom protocol-specific fuzzing, in terms of number of states and other graph complexity metrics.

\vspace{3pt}
%\noindent
\textbf{RQ3: Can \stateafl{} achieve a high fuzzing performance?} The main principle for effective fuzzing is to generate large amounts of inputs over a long period of time. In order to assure that \stateafl{} can perform a high number of tests in the long run, it is important to minimize its overhead on the execution of the server under test. Thus, we evaluate the performance slow-down of the instrumented targets compared to non-instrumented execution.

\vspace{3pt}

%\subsection{Experimental setup}

To assess the feasibility of the approach and to support reproducible experimentation, we implemented \stateafl{} and integrated it with \textsc{ProFuzzBench}, a public benchmark for network fuzzers \citep{profuzzbench2021}. The benchmark includes 13 open-source network servers (Table~\ref{tab:benchmark_targets}). These targets are quite diverse with respect to several aspects: they cover 10 network protocols that have been typical targets of previous fuzzing studies; they are implemented both in C and in C++; they include both TCP and UDP, and both binary and text protocols; they adopt a variety of APIs (e.g., \textsc{send}/\textsc{recv} vs. \textsc{fwrite}/\textsc{fread} for networking, \textsc{pthreads} vs. \textsc{fork} for multiprocessing). \textsc{ProFuzzBench} automates the setup and the execution of the target servers using Docker containers, in a reproducible way. Moreover, \textsc{ProFuzzBench} configures the servers according to the best practices for coverage-driven fuzzing. In particular, the targets are patched to disable sources of randomness (e.g., pseudo-random number generators) in order to have reproducible behavior (i.e., if the program is executed again with the same input, then the same execution path is covered), which is an implicit assumption for coverage-driven fuzzing techniques. 
The experiments adopt the seed inputs from the \textsc{ProFuzzBench} project, where both practitioners and researchers contributed with both benchmark targets and with seeds for these targets. These seeds reflect typical basic usage of the servers according to their experience. The seeds include correct authentication and passwords (otherwise, the fuzzer would waste significant time before getting access to the server), and other frequent commands for the protocol (e.g., for FTP, the seeds get the list of files on the server, create directories and move across them, etc.). \tablename{}~\ref{tab:benchmark_targets} provides the number of unique commands in the seeds for each target server. 
We remark that the need to provide initial seeds for the target server is a problem for any greybox fuzzing approach, in terms of automation and ability to work out-of-the-box for new software to test. We leave this aspect out of the scope of this work.

\begin{table}[!htp]
\centering
\caption{Benchmark targets}
\label{tab:benchmark_targets}
%\vspace{-0.2cm}
{
%\small
\footnotesize
\setlength{\tabcolsep}{4pt}
\begin{tabular}{lllllll}\toprule
\textbf{Target} & \textbf{Protocol} & \textbf{Type} & \textbf{Transport} & \textbf{Lang.} & \textbf{Multiproc.} & \textbf{Seeds} \\
\midrule
Bftpd		& FTP & Text & TCP & C & fork & 54 \\
Dcmtk	& DICOM & Binary & TCP & C++ & pthreads & 4 \\
Dnsmasq		& DNS & Binary & UDP & C & fork & 9 \\
Exim		& SMTP & Text & TCP & C & fork & 9 \\
Forked-daapd	& DAAP & Text & TCP & C & pthreads & 65 \\
Kamailio	& SIP & Text & UDP & C & fork & 3 \\
LightFTP	& FTP & Text & TCP & C & pthreads & 10 \\
Live555		& RTSP & Text & TCP & C++ & N/A & 33 \\
OpenSSH		& SSH & Binary & TCP & C & fork & 22 \\
OpenSSL		& TLS & Binary & TCP & C & N/A & 8 \\
ProFTPD		& FTP & Text & TCP & C & fork & 54 \\
Pure-FTPd	& FTP & Text & TCP & C & fork & 54 \\
TinyDTLS	& DTLS & Binary & UDP & C & N/A & 5 \\
\bottomrule
\end{tabular}
}
\end{table}

The experimental evaluation compares \stateafl{} with two baseline fuzzers. The baseline fuzzers were selected such that: (i) They are not limited to specific network protocols, but are applicable to a large set of network targets, including the ones in \textsc{ProFuzzBench}; this leaves out fuzzers that are highly-customized for a specific protocol (e.g., TLS \citep{de2015protocol}, DTLS \citep{fiterau2020analysis}) but are not applicable to other protocols; (ii) They adopt state-of-the-art greybox, coverage-driven techniques, in order to evaluate how the proposed greybox solution relates to them. The two baseline fuzzers are:

\begin{itemize}
  \item \aflnwe{}: It is a ``network-enabled'' version of \afl{}, with minor changes to send mutated inputs over a TCP/IP socket instead of using file I/O. It adopts the same mutation operators and coverage analysis from \afl{}.
  \item \aflnet{}: It is another fork of \afl{}, with extensive modifications for stateful network fuzzing. It organizes an input as a session of multiple messages, and adds mutation operators at the message level (e.g., dropping or duplicating individual messages, rather than bytes or blocks). Moreover, it relates each input message to a protocol state reached by that message, where the protocol state is represented by the ``status'' code from the response by the server.
\end{itemize}

These two tools represent different points in the design space of greybox network fuzzers. On the one hand, \aflnwe{} is a pure greybox, coverage-driven fuzzer, and it is a baseline to evaluate the relative merit of stateful fuzzing compared to plain coverage-driven fuzzing. On the other hand, \aflnet{} is a stateful network fuzzer that performs protocol state inference. Differently from the proposed \stateafl{} fuzzer, \aflnet{} relies on the contents of response messages to infer protocol states. Therefore, to be applicable, \aflnet{} must be customized with protocol-specific parsers, in order to extract status codes from the messages (where available). Therefore, \aflnet{} comes with parsers for a set of common protocols, and \textsc{ProFuzzBench} extended \aflnet{} with more parsers to support the network servers under test (Table~\ref{tab:benchmark_targets}). For some protocols (e.g., TinyDTLS), response messages do not have status codes; thus, the protocol parsers generate status codes from other fields (e.g., in DTLS, by joining the \emph{content type} field from the header, and the \emph{message type} field from the payload), based on protocol knowledge of \aflnet{}'s developers.

\stateafl{} overcomes the need for protocol-specific parsers, by instrumenting the target process and analyzing its memory at run-time, in order to be more broadly applicable without the need for manual customizations. In our evaluation, we analyze whether the protocol state inference by \stateafl{} can overcome the lack of protocol parsers. 
The experimental plan consists of a total of 156 experiments, with 4 repeated fuzzing experiments for each of the 13 target servers and of the 3 fuzzers, over a period of 24 hours for each experiment. We execute experiments on the Google Cloud Platform, using E2 high-memory VM instances with 4 vCPUs, with a dedicated vCPU for each replication.

In our evaluation, the \stateafl{} fuzzer could successfully run on all of the target servers, without any protocol customization. \stateafl{} automatically instruments I/O APIs from the standard C library, to trigger \textsc{on\_send} (on \textsc{send}, \textsc{sendto}, \textsc{sendmsg}, \textsc{write}, \textsc{fprintf}, and \textsc{fwrite}) and to trigger \textsc{on\_receive} (\textsc{recv}, \textsc{recvfrom}, \textsc{recvmsg}, \textsc{read}, \textsc{fgets}, \textsc{fread}). In only one of the targets (Forked-daapd), which performs network I/O through the \emph{libevent} API, we needed to configure \stateafl{} to probe the \textsc{evhttp\_request\_get\_uri} and \textsc{evhttp\_send\_reply} API functions, in order to keep track of its request/reply loop. The information about the names of these APIs is easily available to developers, and can be learned from a quick inspection of the target server. No modification of \stateafl{} was needed, as it instruments these APIs in the same way of other APIs.

\section{Experimental results}
\label{sec:evaluation_results}

\subsection{Coverage and vulnerabilities}
\label{subsec:cov_vuln}

\begin{figure*}[!ht]
  \begin{center}
  \includegraphics[width=\columnwidth]{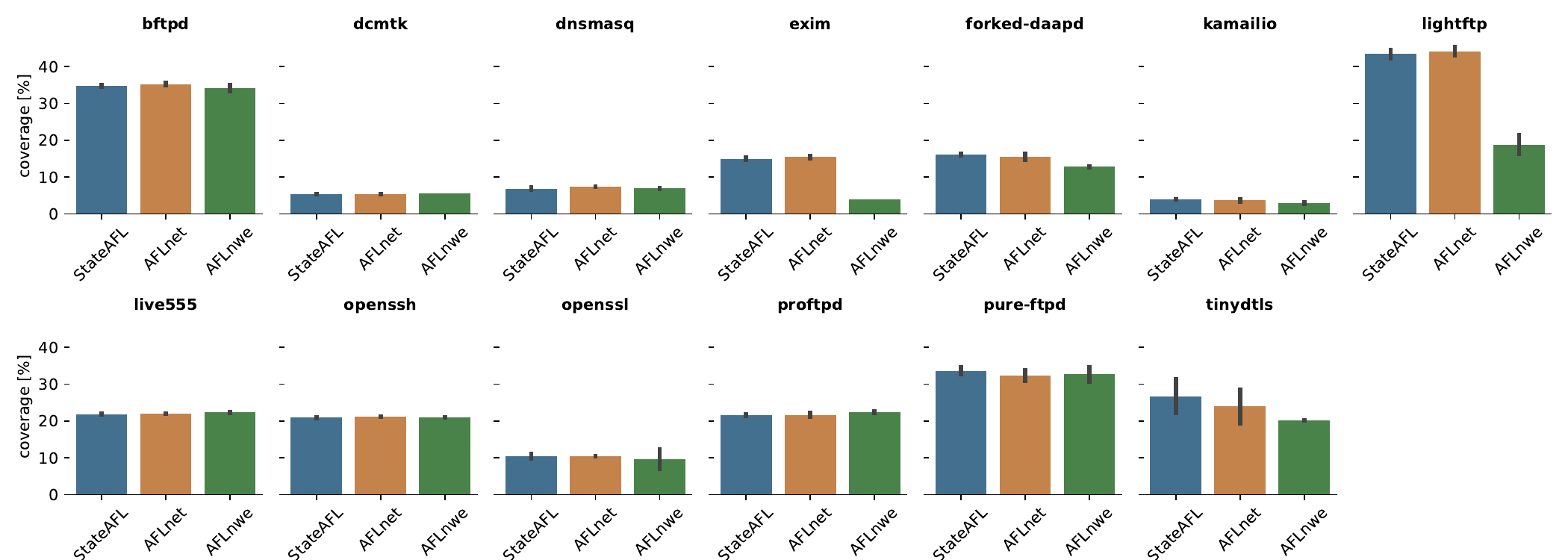}
  %\vspace{-0.2cm}
  \caption{Edge coverage after 24 hours of fuzzing.}
  \label{fig:coverage}
  \end{center}
\end{figure*}

In \tablename{}~\ref{tab:crashes} and \figurename{}~\ref{fig:coverage}, we report respectively on crashes and on coverage for each target and for each fuzzer, after 24 hours of fuzzing. 
For coverage, we focus on edge coverage and do not show line coverage for the sake of space, as it exhibits similar results to edge coverage.

For coverage (\figurename{}~\ref{fig:coverage}), we have different outcomes depending on the target. For 6 targets (Bftpd, Dcmtk, Dnsmasq, Live555, OpenSSH, ProFTPD), we notice that all of the fuzzers achieved a similar coverage. In the other cases, the stateful fuzzers (\stateafl{} and \aflnet{}) achieved higher coverage than the non-stateful \aflnwe{} fuzzer. In particular, for three targets (LightFTP, Exim, TinyDTLS), the gap is larger, while for the remaining targets (OpenSSL, Pure-FTPd, Forked-daapd, Kamailio), the gap is relatively small, but there is a higher variability between experimental runs.

In the 6 targets with similar coverage, both of the stateful fuzzers could not increase coverage compared to plain greybox fuzzing. A possible reason is that the server behavior is only weakly correlated to the current state of the process, as it is influenced mostly by the current input. Therefore, stateless fuzzing could eventually catch up with the stateful fuzzers over the course of the experiment. 
In the other 7 cases, the stateful fuzzer benefited from inferring protocol states. When a message succeeds at discovering a new state, it uses the state (and the messages sent up to that point) as a starting point to generate more inputs. For example, the fuzzer can add further messages after that starting point, and cover new parts that are enabled by the current protocol state. Instead, a stateless fuzzer does not reason in terms of sequences of states, and focuses on mutating the ``interesting bits'' of the input that recently changed the coverage, slowing down the effectiveness of fuzzing. This is further discussed in the analysis of the next subsection (subsec.~\ref{subsec:state_inference}).

The availability of ``status codes'' in the response also seems to have an influence, as in the case of TinyDTLS and, to a minor degree, in the case of OpenSSL. These projects implement binary protocols, and lack a ``status'' code in the response message. For these protocols, the custom parsers in \aflnet{} produce surrogate values, which are computed from other fields in the header and payload. Therefore, the server gives a weaker indication about the current protocol state. In \stateafl{}, the (un)availability of status code in the response does not affect the fuzzing process, as the protocol state is automatically inferred from the analysis of the memory of the target server.

\begin{table}[!htp]
\centering
\caption{Unique bugs found by the fuzzers after 24 hours of fuzzing. The \checkmark{} denotes unique bugs, and the numbers in parentheses denote average non-deduplicated crashes.}
\label{tab:crashes}
%\vspace{-0.2cm}
{
%\small
\footnotesize
\setlength{\tabcolsep}{4pt}
\begin{tabular}{llll}\toprule
\textbf{Target} & \textbf{\aflnwe{}} & \textbf{\aflnet{}} & \textbf{\stateafl{}}  \\
\midrule
Dcmtk		& \checkmark{} (1) & \checkmark{} (10) & \checkmark{} (9) \\
Dnsmasq		& \checkmark{} (54) & \checkmark{} (57) & \checkmark{} (66) \\
Live555		& \checkmark{} (175) & \checkmark{} (211) & \checkmark{} (187) \\
ProFTPD		& --- & --- & \checkmark{} (1,051) \\
TinyDTLS		& \checkmark{} (20) & \checkmark{} (37) & \checkmark{} (56) \\
\bottomrule
\end{tabular}
}
\end{table}

For bugs, we focus on identifying which are the targets where a fuzzer found any crash. We do not consider the absolute number of crashes reported by the fuzzers (``unique crashes''), which is widely acknowledged as an unreliable metric, since the crashes are duplicates of the same underlying vulnerability. For example, the ``unique crashes'' terminology has recently been dropped by the community working on AFL-based fuzzers \citep{aflplusplus2021rename}. %https://twitter.com/aflplusplus/status/1468942228844433413
Therefore, to deduplicate bugs (i.e., to identify crashes that are caused by the same root cause), we first grouped the crashes with respect to their call stack at the time of the crash. Then, we manually analyzed the call stacks, in order to establish whether two groups of crashes with different call stacks where still due to the same bug. Even if different groups of crashes had small differences in the call stack, they were still accounted to the same unique bug due to their semantic similarity. For example, for Live555, two different call stacks with \emph{handleCmd\_SETUP} are both related to the \emph{SETUP} command in the RTSP protocol, and were considered as effect of the same bug.

\tablename{}~\ref{tab:crashes} provides information on the targets that crashed in the experiments. The checkmark symbol ``\checkmark{}'' denotes a unique bug found by fuzzing (in these experiments, at most one for each target listed in the table). 
\stateafl{} was able to crash the same targets of both \aflnwe{} and \aflnet{} (Dcmtk, Dnsmasq, Live555, TinyDTLS), but without relying on protocol customizations. Moreover, \stateafl{} was the only fuzzer able to find crash-inducing inputs for the ProFTPD target. The other fuzzers did not find any bug not  found by \stateafl{}. 
The underlying bug is a heap buffer over-read, which could not be triggered by the other fuzzers since it is difficult to reproduce. ProFTPD introduces its own heap memory allocator on top of the GNU C library, by allocating memory blocks from pre-allocated pools. Therefore, the buffer overrun could not be reliably detected since the buffer overrun can still access to valid memory areas in the same pre-allocated pool. In the case of \stateafl{}, the bug was triggered since the fuzzer put more stress on the memory allocator, thus fragmenting the data and making the buffer over-run more likely to access to invalid memory areas.

All of these crashes were found within one hour of fuzzing. For the other targets, none of the fuzzers were able to find deeper vulnerabilities within 24 hours. However, as discussed later (subsec.~\ref{subsec:state_inference}), \stateafl{} is able to restrict the set of inferred protocol states, which contributes to find deep bugs in the long run, by avoiding repeating the same tests on redundant states. For example, inferring two redundant states (i.e., the server does not actually exhibit different behaviors in these states) leads the fuzzer to repeatedly apply the same fuzz input on both the two states, causing a waste of computational efforts. 
We remark that the main contribution is the increase in automation (as it avoids the user to write protocol-custom code) and in broadening the scope of fuzzing (as it can support more protocol types with lower effort), while keeping a performance level comparable to existing fuzzers in terms of coverage and bugs found.

\subsection{Protocol state inference}
\label{subsec:state_inference}

\begin{figure}[!ht]
  \begin{center}
  \includegraphics[width=0.8\columnwidth]{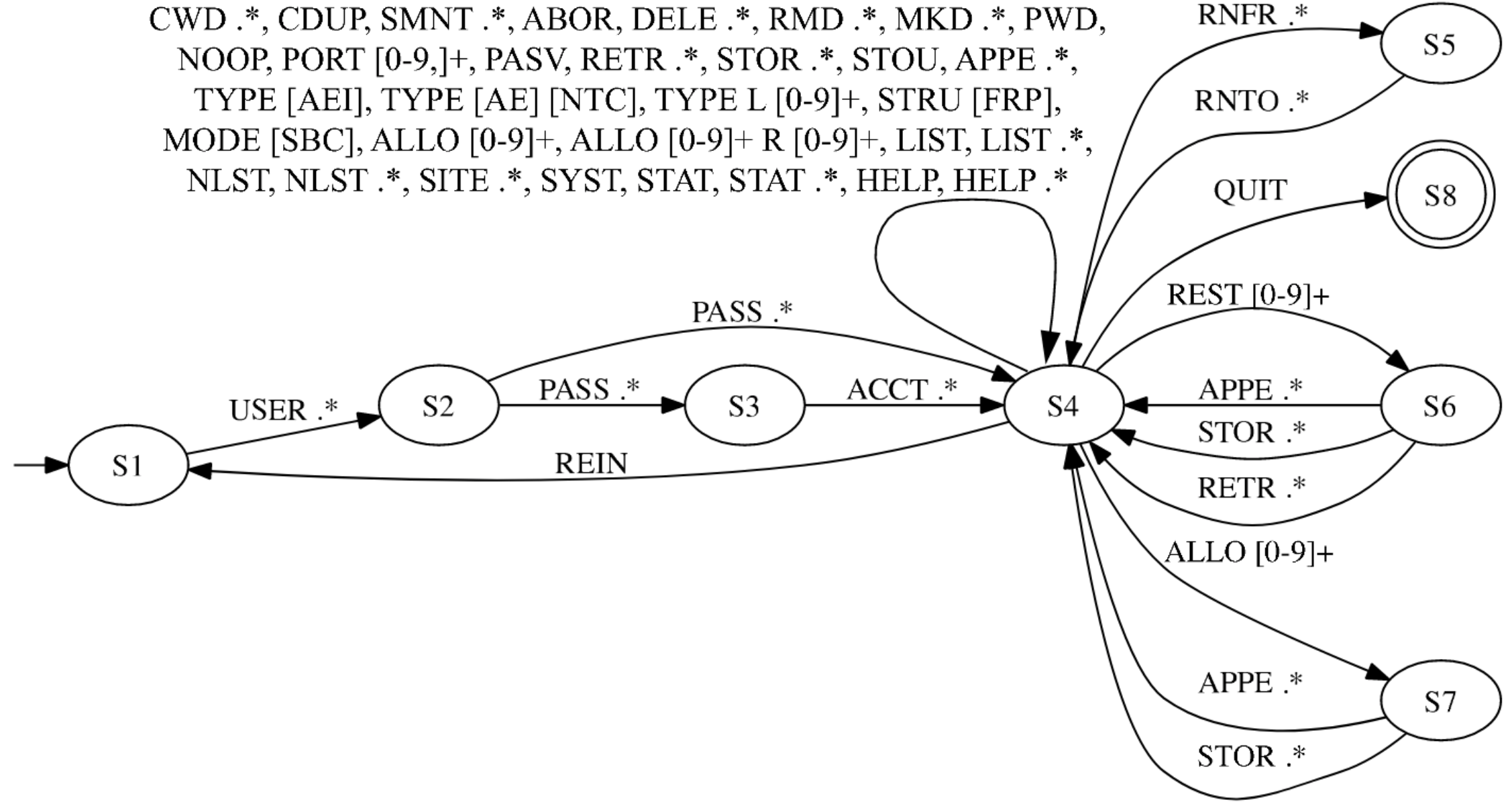}
  \end{center}
  \caption{Reference state machine for the FTP protocol, by \citet{antunes2011automatically}.}
  \label{fig:ftp-ground-truth}
\end{figure}

To get a better understanding of the protocol state machine inferred by \stateafl{}, we first analyze one of the targets in a qualitative way. We focus on the FTP protocol, since it is a plain-text protocol that is simple to be manually interpreted, and that has been targeted by many fuzzing studies \citep{profuzzbench2021}. Moreover, the FTP protocol is simple enough to be modeled with a small state machine derived from the protocol specification, which can serve as a \emph{ground truth} for our analysis. In particular, we consider as reference the model by \citet{antunes2011automatically} based on the RFC 959, shown in \figurename{}~\ref{fig:ftp-ground-truth}. The initial states of the protocol represent the phases of user authentication ($S1$, $S2$, $S3$); most of the commands do not affect the protocol state (e.g., reading or updating the configuration of the server), as they leave the server in state $S4$; only a small set of commands (e.g., REST for resuming a transfer) introduce additional states.

Among the four FTP servers from the benchmark, we focus on LightFTP. In order to interpret the state machine inferred by \stateafl{}, we manually analyze both the source code of the server and the inputs that covered the states. This implementation of FTP is the simplest one and amenable for our manual analysis, as the core of the protocol implementation (not considering the parsing of the configuration file and of the command line) consists of about 1.7 kLoCs, and is limited to only one source file and one header file (\texttt{ftpserv.c} and \texttt{ftpserv.h}), thus mitigating the risk of an incorrect manual interpretation. The long-lived state of the server is all included within one data structure (\texttt{FTPCONTEXT}) allocated on the stack, which contains the socket handles, IP port numbers and addresses for the client and the server, the currently-opened file and access mode, the current working directory, and handles for a worker thread and for a mutex. Among the FTP commands implemented in this server, only few ones update the state of the server. Differing from the state machine based on the standard (\figurename{}~\ref{fig:ftp-ground-truth}), some commands have side effects on long-lived data, since they launch a worker thread to access the file system and to transfer data back to the client through a separate connection (e.g., the LIST and MLSD commands for listing the contents of a folder; the STOR, RETR, and APPE commands for file transfer); other commands, such as PORT, PASV, EPSV, and PBSZ change the configuration of the server (e.g., the client port to be used for the data connection).

\begin{figure}[!ht]
  \begin{center}
  \includegraphics[width=0.5\columnwidth]{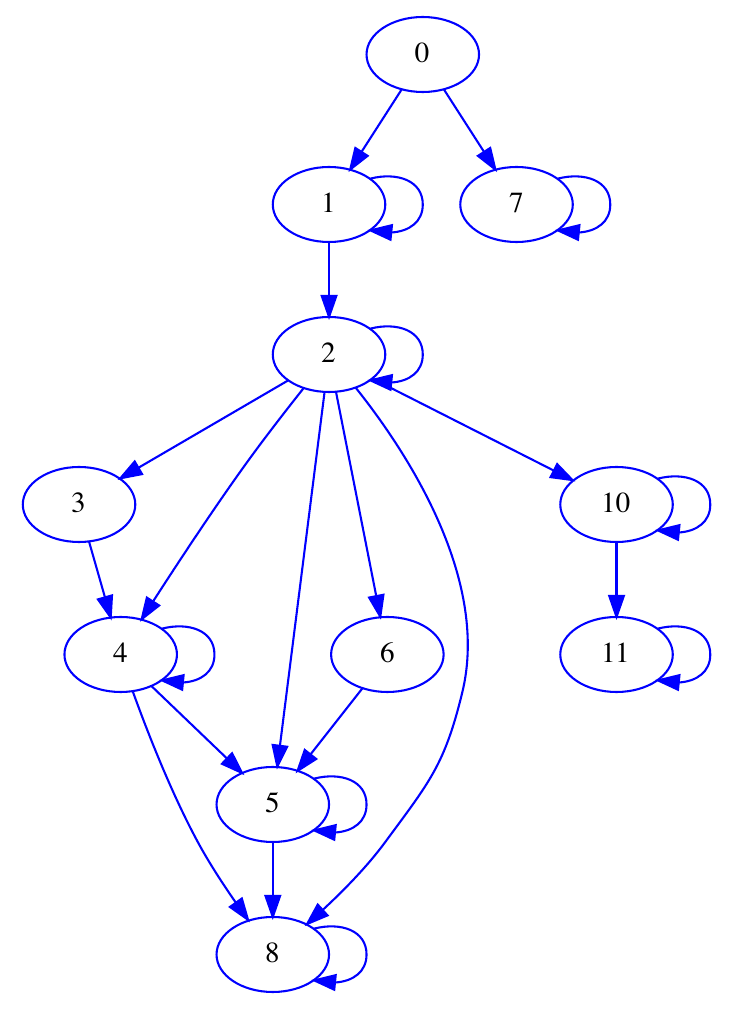}
  \end{center}
  \caption{State machine for LightFTP inferred by \stateafl{} after 24 hours.}
  \label{fig:lightftp-ipsm-full}
\end{figure}

\figurename{}~\ref{fig:lightftp-ipsm-full} shows a protocol state machine inferred by \stateafl{} for the LightFTP  server. The state machine starts from a ``dummy'' initial state $0$; the other states represent unique in-memory states of long-lived data in the target server, identified by an incremental number; an edge represents a request/reply pair between the server and the client; the edges can be self-transitions in the same state, which is the case of messages without side effects (e.g., read-only operations); or, the edges can bring the server to a different state, which reflect changes in long-lived data. As in the reference model, the states $0$, $1$ and $2$ are followed during user authentication. In the case of unsuccessful authentication, the state machines moves to state $7$. Most of the commands are stateless as in the reference model, and are represented by the self-transition in state $2$. In this case, \stateafl{} correctly recognizes that the server is not changing state from the analysis of process memory, thus avoiding to add more states when stateless commands are issued. The server moves to the other states in the case of the PORT command (states $3$, $6$, and $10$) and the LIST command (states $4$, $5$, $8$, $11$). Ideally, the state machine should use  fewer states to represent the conditions that the data connection has been configured (PORT) and that the worker thread has been launched (LIST). However, the contents of the long-lived data vary across different executions of these commands, since the data depend on the parameters of the PORT command, and on non-determinism in the initialization of the worker thread. \stateafl{} performs locality-sensitive hashing to cluster these different contents of the long-lived data into few states of the inferred state machine, thus limiting the growth of redundant states.

\begin{figure}[!ht]
  \begin{center}
  \subfloat[\stateafl{}.\label{fig:lightftp-ipsm-seed-stateafl}]{%
        \raisebox{0mm}{\includegraphics[width=0.6\columnwidth]{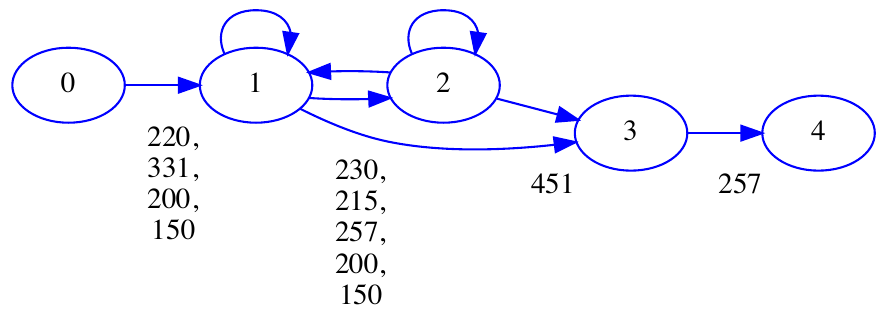}}%
  }
  \qquad
  \subfloat[\aflnet{}.\label{fig:lightftp-ipsm-seed-aflnet}]{%
        \includegraphics[width=\columnwidth]{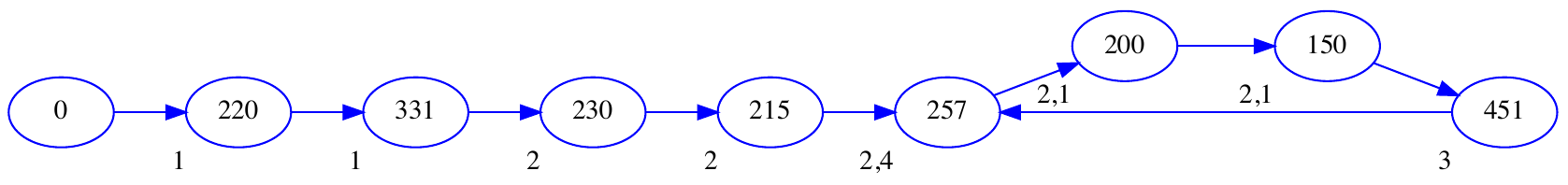}%
  }
  \end{center}
  \caption{Inferred state machine for LightFTP using two seed inputs. States are annotated with the corresponding states for the other fuzzer.}
  \label{fig:lightftp-ipsm-seed}
\end{figure}

To get more insights about the state machines inferred by \aflnet{} and \stateafl{}, we analyzed the overlap between their state machines for the LightFTP server. \figurename{}~\ref{fig:lightftp-ipsm-seed-stateafl} and \figurename{}~\ref{fig:lightftp-ipsm-seed-aflnet} show respectively the state machine for \stateafl{} and \aflnet{}, by running the same two seed inputs. These inputs establish a session by logging-in, and perform basic operations such as listing files in the root folder, querying the OS version of the server, setting the data connection, creating a folder, and quitting the session. In each figure, the states inferred by one fuzzer are annotated (nearby the vertex of the graph) with the label of the corresponding state of the other state machine. The set of status codes returned by the server (i.e., $220$, $331$, etc., used by \aflnet{}, in \figurename{}~\ref{fig:lightftp-ipsm-seed-aflnet}) is larger than the set of unique in-memory states reached by the target server (i.e., $1$, $2$, $3$, and $4$ inferred by \stateafl{}, in \figurename{}~\ref{fig:lightftp-ipsm-seed-stateafl}). This overlap highlights an important difference between status codes and the concept of ``state''. The status code only reflects the outcome of the most recent command (i.e., the latest request/reply iteration), regardless of which operations were previously performed on the server, which side effects have (or have not) accumulated within the target process, and how the server will behave in response to future commands. In this example, several commands return different status codes but do not have side effects on long-lived data of the server, such as the self-transitions in states $1$ and $2$ in \figurename{}~\ref{fig:lightftp-ipsm-seed-stateafl}. The additional states inferred by \aflnet{} are redundant for stateful fuzzing, since applying the same fuzzed message starting from any of the redundant states (e.g., $230$, $215$, $257$, $200$, and $150$ in \figurename{}~\ref{fig:lightftp-ipsm-seed-aflnet}) results in the same behavior of the server, since the in-memory state of the process is always the same. In turn, these additional states result in wasted attempts by \aflnet{} to fuzz the server under (apparently) different conditions.

\begin{table}[t]
\centering
\caption{Metrics about the inferred protocol state machines.}
\label{tab:graph_stats}
{
%\footnotesize
\scriptsize
\setlength{\tabcolsep}{3pt}
\begin{tabular}{cccccccccc}
\toprule
\textbf{target} & \textbf{fuzzer} & \textbf{vertexes} & \textbf{edges} & \textbf{\makecell{longest dist.\\ from root}} & \textbf{\makecell{out \\ degree}} & \textbf{circuits} \\
\midrule
\multirow{2}{*}{Bftpd} & \aflnet{} & 24 & 183 & 4 & 7 & >1M \\
 & \stateafl{} & 3 & 6 & 1 & 2 & 3 \\
\midrule
\multirow{2}{*}{Dcmtk} & \aflnet{} & 4 & 3 & 1 & 1 & 1 \\
 & \stateafl{} & 15 & 31 & 1 & 1 & 13 \\
\midrule
\multirow{2}{*}{Dnsmasq} & \aflnet{} & 88 & 278 & 5 & 3 & >1M \\
 & \stateafl{} & 52 & 145 & 5 & 2 & >20K \\
\midrule
\multirow{2}{*}{Exim} & \aflnet{} & 12 & 57 & 3 & 4 & 353 \\
 & \stateafl{} & 21 & 45 & 3 & 2 & 19 \\
\midrule
\multirow{2}{*}{Forked-daapd} & \aflnet{} & 7 & 19 & 2 & 2 & 6 \\
 & \stateafl{} & 4 & 4 & 1 & 1 & 1 \\
\midrule
\multirow{2}{*}{Kamailio} & \aflnet{} & 13 & 105 & 1 & 7 & >315K \\
 & \stateafl{} & 2 & 2 & 1 & 1 & 1 \\
\midrule
\multirow{2}{*}{LightFTP} & \aflnet{} & 23 & 176 & 3 & 7 & >1M \\
 & \stateafl{} & 11 & 26 & 3 & 2 & 17 \\
\midrule
\multirow{2}{*}{Live555} & \aflnet{} & 10 & 75 & 2 & 7 & >37K \\
 & \stateafl{} & 16 & 31 & 2 & 1 & 12 \\
\midrule
\multirow{2}{*}{OpenSSH} & \aflnet{} & 111 & 246 & 8 & 2 & >250K \\
 & \stateafl{} & 153 & 467 & 4 & 3 & >1M \\
\midrule
\multirow{2}{*}{OpenSSL} & \aflnet{} & 17 & 26 & 4 & 1 & 5 \\
 & \stateafl{} & 2 & 2 & 1 & 1 & 1 \\
\midrule
\multirow{2}{*}{ProFTPD} & \aflnet{} & 26 & 241 & 4 & 9 & >1M \\
 & \stateafl{} & 4 & 9 & 1 & 2 & 5 \\
\midrule
\multirow{2}{*}{Pure-FTPd} & \aflnet{} & 29 & 294 & 4 & 10 & >1M \\
 & \stateafl{} & 10 & 29 & 1 & 2 & 15 \\
\midrule
\multirow{2}{*}{TinyDTLS} & \aflnet{} & 7 & 19 & 2 & 2 & 15 \\
 & \stateafl{} & 9 & 18 & 1 & 1 & 6 \\
\bottomrule
\end{tabular}
}
\end{table}

\tablename{}~\ref{tab:graph_stats} provides statistics about the inferred protocol state machines, for both \aflnet{} and \stateafl{}, over all of the 13 target servers. The values in the table are the mean across repetitions. In the case of \aflnet{}, the vertexes represent the ``status'' code returned by the server in a request/reply iteration, while in \stateafl{} the vertexes represent unique memory states. The number of states for \stateafl{} is lower than \aflnet{} for almost all of the targets. Therefore, the other metrics in \tablename{}~\ref{tab:graph_stats} also tend to be lower for \stateafl{} (number of edges; longest distance between the root node and other nodes; the degree of output transitions from a state; the number of circuits). \stateafl{} infers recurring states for most of the protocols, as for LightFTP in \figurename{}~\ref{fig:lightftp-ipsm-seed-stateafl}.

\aflnet{} and \stateafl{} inferred different protocol state machines from the four FTP servers (LightFTP, Bftpd, Pure-FTPd, ProFTPD). Despite these servers implement the same protocol, it is typical for different implementations to cover a different subset of the protocol specification, or to include extensions from later standards or from the vendor \citep{antunes2011automatically,poll2015protocol}. In all cases, the state machines inferred by \stateafl{} have a lower number of states and closer to the reference model of \figurename{}~\ref{fig:ftp-ground-truth}.

In two cases (Kamailio and OpenSSL), the protocol state machine inferred by \stateafl{} consists of only two states, including the dummy state $0$ and only one, fixed state over the course of the session. In the case of Kamailio, the default configuration only performs stateless routing of SIP requests, which are only based on the contents of the request; stateful processing needs to be enabled by configuring an optional module, as it is more resource-demanding and aimed for advanced use cases \citep{nick2019stateless}. By analyzing the dumps collected by \stateafl{}, we found that long-lived data indeed do not change across iterations for these servers. In the case of OpenSSL, the fuzzer could not identify new states, since it is a highly-structured binary protocol, for which it is difficult to generate new sequences of valid messages. This is a general limitation of mutation-based fuzzers, that could be addressed by means of structure-aware fuzzing techniques \citep{google2017structure}. In both cases, \stateafl{} can detect that the inputs hit the same state, thus avoiding to perform redundant tests.

\subsection{Performance}
\label{subsec:performance}

\begin{figure*}[!htb]
  \begin{center}
  \includegraphics[width=\columnwidth]{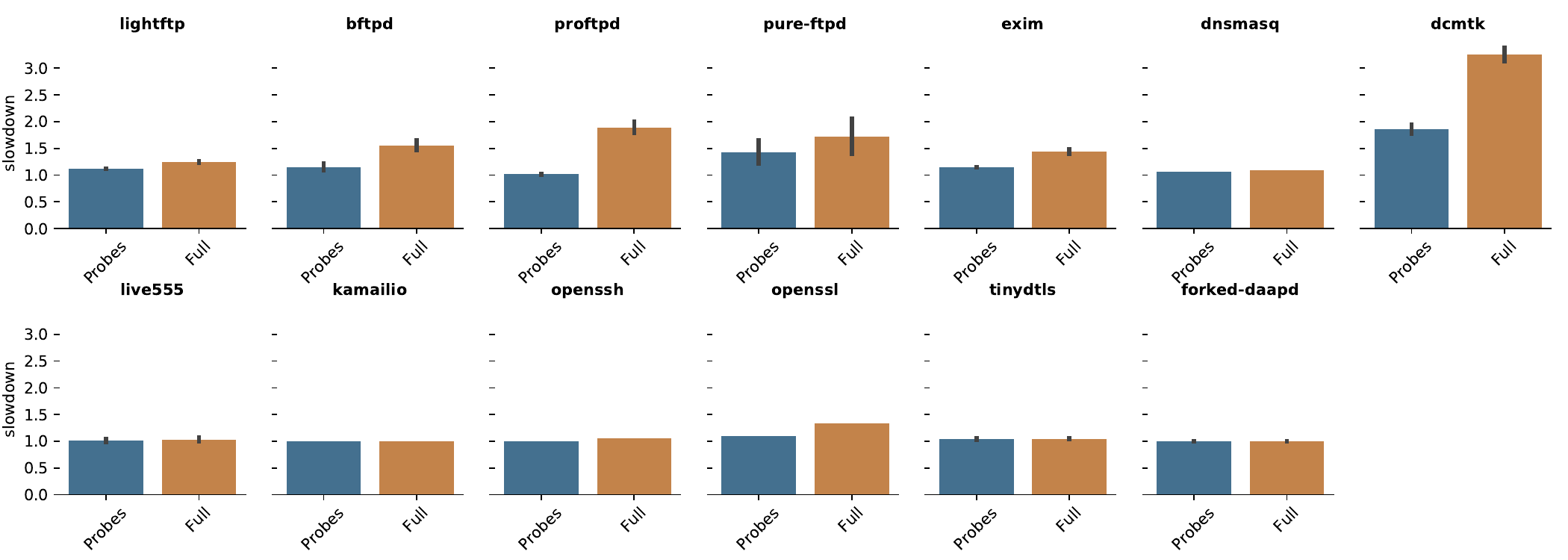}
  %\vspace{-0.2cm}
  \caption{Slowdown of execution time under \stateafl{}, respectively when the target is only instrumented with callbacks for data collection (\emph{probes}), and when the instrumentation also performs post-execution analysis (\emph{full}). The slowdown is normalized with respect to execution without instrumentation.}
  \label{fig:slowdown}
  \end{center}
\end{figure*}

Finally, we evaluated the performance overhead of the instrumentation code injected by \stateafl{} in the target process. \figurename{}~\ref{fig:slowdown} reports the execution time of the target servers when running seed inputs. The execution time under \stateafl{} is normalized with respect to the execution time without instrumentation (e.g., a 1.1x slowdown means that the execution takes 10\% more time to complete). The instrumentation code mainly consists of: (i) the probes injected where the target server allocates memory and performs network I/O, to make callbacks for data collection (Algorithms \ref{alg:on_receive} to \ref{alg:dump_current_state}); (ii) post-execution analysis (Algorithms \ref{alg:on_process_end} and \ref{alg:save_state_seq}). Therefore, we separately evaluate the impact of these two types of instrumentation code.

\figurename{}~\ref{fig:slowdown} shows the slowdown respectively when only the probes are injected without any post-execution analysis (labeled with \emph{Probes}), and when the instrumentation code also includes the post-execution analysis (labeled with \emph{Full}). For some targets, the relative slowdown is negligible (i.e., close to 1x). The relative slowdown is noticeable for those target servers that take less time to execute the inputs, and that allocate a larger amount of long-lived data to be analyzed. In these cases, the slowdown was around 1.5x the execution time of the non-instrumented server, and around 3x in the worst case of Dcmtk. In these cases, the non-instrumented execution time takes less than 100ms to process the inputs. Most of the slowdown comes from the post-execution analysis, which computes hashes from memory snapshots. This analysis takes fractions of ms in the best cases, and around 100ms in the worst cases. Instead, for those targets that take longer to process the inputs (e.g., Forked-daapd), the relative weight of the post-execution analysis becomes negligible. Moreover, the slowdown caused by \stateafl{} is balanced by a reduction of redundant states in the inferred state machines, leading to less states to be explored by fuzzing and less wasted inputs, thus achieving similar or better code coverage.

\begin{figure*}[!htb]
  \begin{center}
  \includegraphics[width=\columnwidth]{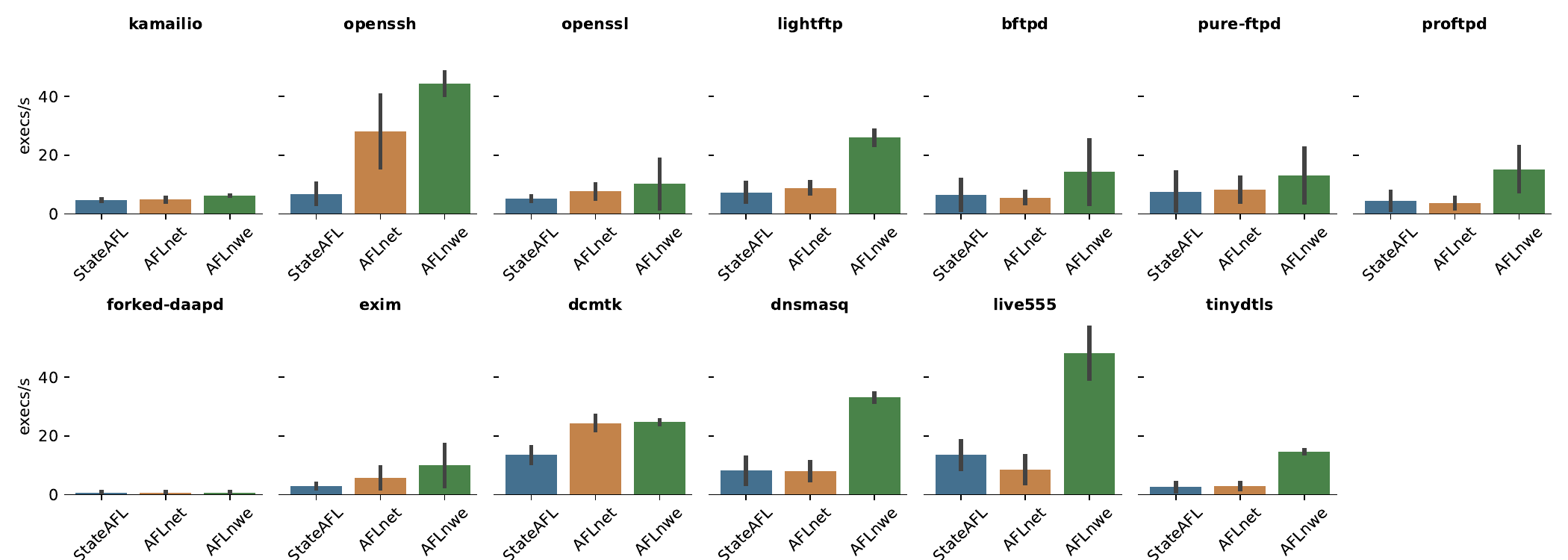}
  %\vspace{-0.2cm}
  \caption{Fuzzing throughput (executions of the target server per second), averaged over 24 hours of fuzzing and 4 repetitions.}
  \label{fig:fuzz-throughput}
  \end{center}
\end{figure*}

Finally, \figurename{}~\ref{fig:fuzz-throughput} reports the throughput of the fuzzers, in terms of executions of the target server per second, averaged over 24 hours of fuzzing and 4 repetitions. The \aflnwe{} achieved the highest throughput across most of the benchmark targets. Compared to the other two fuzzers, \aflnwe{} is \emph{not} a message-oriented fuzzer, as it sends fuzz inputs as an uninterrupted stream of bytes. Instead, \aflnet{} and \stateafl{} are message-oriented fuzzers, which alternate between sending request messages and receiving (and analyzing) response messages, which introduces short delays. \aflnet{} and \stateafl{} achieved a comparable fuzzing throughput for most of the benchmark targets. For few targets (OpenSSH and Dcmtk), \stateafl{} exhibited a significantly lower fuzzing throughput than \aflnet{}. In the case of Dcmtk, we can attribute this gap to the combined effect of short absolute execution time of the target, and the higher slowdown caused by the instrumentation (\figurename{}~\ref{fig:slowdown}). For OpenSSH, the lower throughput was caused by an additional delay between request messages, which was configured to make the analysis of in-memory states more deterministic. A potential extennsion to avoid the need for such delays, and in general for improving the fuzzing throughput, is represented by ongoing research on \emph{snapshot-based fuzzing}, which saves and restores the state of the entire server process at selected times \citep{li2022snpsfuzzer,andronidis2022snapfuzz}. Please note that snapshot-based fuzzing is complementary area of research to \stateafl{}, which infers states from a fine-grained analysis of process memory, and would guide the snapshot-based process by identifying unique application-level states.

\section{Conclusion}
\label{sec:conclusion}

This paper presented \stateafl{}, a coverage-driven fuzzer for stateful network servers. We designed the fuzzer to not rely on manual customizations for the protocol under test, in order to make stateful fuzzing more broadly applicable. The fuzzer leverages compile-time instrumentation to insert probes, which take snapshots of long-lived data at each protocol iteration. Then, the fuzzer uses fuzzy hashing to map the snapshots to a unique state identifier, in order to infer protocol states.

We implemented and released \stateafl{} as open-source software, and experimentally evaluated it on a benchmark of network servers. The experimental evaluation showed that \stateafl{} can match a protocol-custom fuzzer in terms of both code coverage and vulnerabilities, and can even exceed it for some targets. Moreover, \stateafl{} only introduces a limited overhead on the execution of the server under test. We also presented a qualitative analysis of the states inferred by \stateafl{}. The qualitative analysis pointed out an important insight about stateful fuzzing: the response codes returned by many protocols are often not representative of the current state of the server, but only reflect the outcome of the last request. For this reason, inferring states for response codes can significantly inflate the protocol state machine, leading to redundant fuzz tests. Having knowledge about the actual state of the server can be exploited by the fuzzer to avoid the redundant tests.

We expect that future work on stateful network protocol fuzzing will develop new solutions based on this observation. Potential directions for future research include new solutions from inferring states from memory analysis, such as by using techniques for static and dynamic program analysis, and new heuristics for generating fuzz inputs tailored for stateful protocols, such as algorithms for selecting which protocol states to fuzz and which portions of the input to mutate. 
Early work on these areas include \citet{ba2022stateful} and \citet{li2022snpsfuzzer} for efficient definition of states based on program analysis, and \citet{liu2021state} on state selection algorithms. 
We also leave to future work the application of the proposed approach to binary-only programs. Since the instrumentation is limited to identifying and changing calls to library APIs, without changing the control and data flow, binary rewriting techniques represent good candidates for further research on this aspect \citep{dinesh2020retrowrite,duck2020binary}. 
Finally, we expect future work to explore more applications of stateful fuzzing beyond network protocols, such as for the security testing of local stateful applications based on inter-process communication.

\begin{acknowledgements}
I am grateful to Van-Thuan Pham (University of Melbourne) for the constructive discussions and the encouragement during this work. This work has been partially supported by the Google Cloud research credits program, and by the FRA programme (project OSTAGE) at Università degli Studi di Napoli Federico II.
\end{acknowledgements}

\bibliographystyle{spbasic}      % basic style, author-year citations
\bibliography{references}

\end{document}